\documentclass[natbib]{cta-author}
\usepackage{natbib}
\usepackage{float}
\usepackage{graphicx} %[dvips]
\usepackage{tikz}
\usepackage{pgfplots}
\usepackage{color}
\usepackage{xcolor}
\usepackage{amstext,epsfig,amssymb,amsbsy,amsmath}
\usepackage{mathtools}
\usepackage[output-decimal-marker={.}]{siunitx}
\usepackage{eurosym}
\usepackage{numprint}
\usepackage{microtype}
\usepackage[shortlabels]{enumitem}
\usepackage{array}
\usepackage{tabularx}
\usepackage{multirow}
\usepackage{multicol}
\usepackage{setspace}
\usepackage{verbatimbox}
\usepackage[caption=false,font=footnotesize]{subfig}
\usepackage{wrapfig}
\usepackage{url}
\usepackage{hyperref}
\usepackage{cleveref}

\npthousandsep{'}

\expandafter\def\expandafter\UrlBreaks\expandafter{\UrlBreaks%  save the current one
  \do\a\do\b\do\c\do\d\do\e\do\f\do\g\do\h\do\i\do\j%
  \do\k\do\l\do\m\do\n\do\o\do\p\do\q\do\r\do\s\do\t%
  \do\u\do\v\do\w\do\x\do\y\do\z\do\A\do\B\do\C\do\D%
  \do\E\do\F\do\G\do\H\do\I\do\J\do\K\do\L\do\M\do\N%
  \do\O\do\P\do\Q\do\R\do\S\do\T\do\U\do\V\do\W\do\X%
  \do\Y\do\Z}

{}
{}
{}

\newcommand\Tstrut{\rule{0pt}{2.6ex}}         % = `top' strut
\newcommand\Bstrut{\rule[-0.9ex]{0pt}{0pt}}   % = `bottom' strut

\definecolor{gray1}{rgb}{0.65098,0.65098,0.65098}%
\definecolor{gray2}{rgb}{0.50196,0.50196,0.50196}%
\definecolor{blue1}{rgb}{0.20392,0.30196,0.49412}%
\definecolor{blue2}{rgb}{0.301,0.588,0.733}%
\definecolor{blue3}{rgb}{0.635,0.792,0.886}%
\definecolor{orange1}{rgb}{1.00000,0.60000,0.00000}%
\definecolor{red1}{rgb}{0.75,0,0}%

\pgfplotsset{
    layers/.define layer set={
        background,
        main,
        foreground,
    }{},
    }
\usetikzlibrary{shapes,arrows,calc,positioning,fit,backgrounds,decorations.pathreplacing}
\usepgfplotslibrary{statistics}
\tikzset{
  block/.style = {draw, rectangle,
      minimum height=1cm,
      align = center,
      minimum width={width("Mitigation")+2pt}
  },
  input/.style = {coordinate,node distance=1cm},
  output/.style = {coordinate,node distance=1cm},
  arrow/.style={draw, -latex,node distance=2cm},
  pinstyle/.style = {pin edge={latex-, black,node distance=2cm}},
  sum/.style = {draw, circle, node distance=1cm},
  gain/.style = {
     regular polygon, regular polygon sides=3,
     draw, fill=white, text width=1em,
     inner sep=0mm, outer sep=0mm,
     shape border rotate=-90
  },
  dot/.style={circle,fill,draw,inner sep=0pt,minimum size=3pt}
 }

\begin{document}

%\supertitle{Brief Paper}

% PAPER TITLE
\title{Sharing Reserves through HVDC: Potential Cost Savings in the Nordic Countries}

\author{\au{Andrea~Tosatto$^{1}$},
        \au{Matas~Dijokas$^{1}$},
        \au{Tilman~Weckesser$^{2}$},
        \au{Spyros~Chatzivasileiadis$^{1}$},
        \au{Robert~Eriksson$^{3}$}}

\address{\add{1}{Technical University of Denmark, Department of Electrical Engineering, Kgs. Lyngby,               Denmark}
        \add{2}{Dansk Energi, Grid Technology, Frederiksberg C, Denmark}
        \add{3}{Svenska kraftn\"{a}t, System Development, Sundbyberg, Sweden}
        \add{*}{E-mail: \{antosat,matdij,spchatz\}@elektro.dtu.dk, TWE@danskenergi.dk, Robert.Eriksson@svk.se}}

%% ABSTRACT %%%%%%%%%%%%%%%%%%%%%%%%%%%%%%%%%%%%%%%%%%%%%%%%%%%%%%%%%%%%%%%%%%%%%%%%%%%%%%%%%%%%%%%%%%%%%%%%%%%%%%%%
%\vspace{-1em}
\begin{abstract}
During summer 2018, the Nordic system's kinetic energy dropped below a critical level. As a consequence, Svenska kraftn\"{a}t, the Swedish transmission system operator (TSO), requested the largest production unit to reduce its power output to guarantee system's security. This action resulted in a deviation from the generation dispatch determined by the market and in high costs for the Nordic TSOs. In this regard, this paper presents a tool for comparing mitigation strategies from an economic point of view and evaluates potential economic benefits of utilizing the Emergency Power Control (EPC) functionality of HVDC lines for the provision of fast reserves as a compliment to Frequency Containment Reserves (FCR). Moreover, the analysis is extended to the years 2020 and 2025 using inertia estimations from the Nordic TSOs. The findings of the paper suggest that the frequency of redispatching actions will increase in the future and that the cost of security for Nordic TSOs could be reduced by 70\% if HVDC links are used for frequency support.
\end{abstract}

\maketitle

%% INDEX TERMS %%%%%%%%%%%%%%%%%%%%%%%%%%%%%%%%%%%%%%%%%%%%%%%%%%%%%%%%%%%%%%%%%%%%%%%%%%%%%%%%%%%%%%%%%%%%%%%%%%%%%
%\begin{IEEEkeywords}
%Emergency Power Control, Fast Frequency Reserves, Frequency Containment Reserves, frequency stability, HVDC transmission lines, low inertia, N-1 security, RG Nordic, power redispatch.
%\end{IEEEkeywords}

%% CHAPTERS %%%%%%%%%%%%%%%%%%%%%%%%%%%%%%%%%%%%%%%%%%%%%%%%%%%%%%%%%%%%%%%%%%%%%%%%%%%%%%%%%%%%%%%%%%%%%%%%%%%%%%%%
\section{Introduction}\label{sec:1}
As governments across the world are planning to limit greenhouse gas emissions, the penetration of renewable energy sources has significantly increased in the last years. During the last decade, the total installed wind power capacity has increased from 159 to 651 GW on a global level. In 2019, the global offshore wind industry had a record year with 6.1 GW of new additions. Denmark alone counted as 6\% of the new installed capacity, and forecasts show that investments will not stop here \cite{2_1}. On the one hand, this process represents the first step towards cleaner electricity systems; on the other hand, it causes a shift from synchronous to inverter-based non-synchronous generation, resulting in lower system kinetic energy and reduced power systems robustness to grid disturbances.

Electrical systems are built to continuously match the supply of electricity to customer demand: any mismatch results in a deviation of the frequency from its nominal value (50 Hz in Europe). Small frequency deviations are common during normal operation, mainly caused by load volatility and intermittent renewable generation. To distinguish between normal frequency fluctuation and deviations caused by large imbalances, Transmission System Operators (TSOs) define security thresholds and activate different balancing resources depending on the size of the power deviation. 

In the Nordic region (Denmark, Sweden, Norway and Finland), also referred to as Nordic Synchronous Area (NSA), normal system operation has a standard range of $\pm$100 mHz and Frequency Containment Reserves for Normal operation (FCR-N) are deployed to keep frequency within the normal band \cite{2_2}. When frequency drops below 49.9 Hz, FCR for Disturbances (FCR-D) are activated to mitigate the impact of the disturbance and stabilize the frequency, while Frequency Restoration Reserves (FRR) are used to restore the frequency back to the nominal value. The maximum acceptable Instantaneous Frequency Deviation (IFD) is 1'000 mHz and, in case frequency drops below 48.8 Hz, loads are shed to avoid total system blackout \cite{2_3}. 

The IFD that follows a disturbance depends on the size of the power deviation, on the activation speed of reserves and on the kinetic energy of the system (system inertia). Indeed, kinetic energy stored in the rotating mass of the system opposes changes in frequency after a disturbance and represents the first inherent containment reserve. Due to the replacement of conventional generation with RES, the system's kinetic energy is decreasing, leaving the system more prone to high Rate of Change of Frequency (RoCoF) and larger IFD \cite{2_4}. 

The methodology for calculating the FCR-D requirement consists in a deterministic approach which aims at limiting the frequency deviation to 0.9 Hz after the dimensioning incident, taking into consideration the kinetic energy of the system, the frequency-dependent load and the available FCR-D capacity \cite{2_5}. The considered dimensioning incidents (DI) are the loss of critical components of the system, such as large generators, demand facilities and transmission lines. Since this calculation depends on the kinetic energy of the system, the lowest experienced level of kinetic energy is used, i.e. 120 GWs after the DI. Given the ongoing displacement of synchronous generation, the current requirement might result in insufficient FCR-D dynamic response during low inertia periods, raising concerns among TSOs.

Currently, the dimensioning incident in the NSA is the loss of Oskarshamn 3 (O3), a 1'450 MW nuclear power plant in Sweden (located in the bidding zone SE3) operated by Oskarshamn Kraftgrupp (OKG) \cite{2_6}. The method and the results presented in \cite{2_7} show that, with the current FCR-D requirement, the maximum IFD is exceeded when the Nordic kinetic energy drops below a certain threshold (around 150 GWs with the current dimensioning incident), unless mitigation measures are taken. This has already happened three times in 2018 (June 23-25, July 6-9 and August 11-12) \cite{2_8}. During these three periods, the loss of O3 would have caused an IFD greater than 1'000 mHz, violating the N-1 stability criterion.

To avoid this risk, the Swedish TSO (Svenska kraftn\"{a}t - Svk) ordered OKG to reduce its power output by 100 MW. TSOs are responsible for safe operation of power systems and can give orders to market participants at any market stage (real-time, intra-day, day-1, day-2, day-x) if the system security is in danger. However, this operation comes with high cost, since the affected producers should be compensated for the incurred costs and the substitute power must be procured outside the market operation \cite{2_9}. This mitigation strategy falls in the category of \textit{preventive} actions, which aim at eliminating causes of potentially dangerous situations before these happen \cite{2_19}. The following question arises: are there more cost-efficient options which guarantee safe operation while avoiding expensive redispatching actions?

Decrease of kinetic energy in modern power systems is becoming a common experience of system operators around the world. The first cases date back to 2010 in Ireland, where the Irish system operators (EirGrid and SONI) identified high RoCoF values as potential security problems \cite{2_10a}. In the following years, similar situations were experienced in other small islands, such as Cyprus, Hawaii and New Zealand \cite{2_10b, 2_10c, 2_10d}. More recently, the same problems have reached larger islands, such us Great Britain and Australia, where large part of the electricity demand is supplied by renewable sources \cite{2_10e, 2_10f}. Although large interconnected systems have not experienced such situations yet, many system operators, California Independent System Operator (CAISO) and Electricity Reliability Council of Texas (ERCOT) among others, have performed analyses to calculate the maximum amount of RES their systems can accommodate without experiencing such problems \cite{2_10g, 2_10h, 2_10i}. As more and more system operators are facing these challenges, many technical and regulatory solutions have been proposed in the literature. In \cite{2_10j}, the authors compare different mitigation approaches from a technical point of view, raging from synthetic inertia to more complex control strategies. They also analyze what could be the impact of regulatory measures such as new market constraints, demand side inclusion and adaptive protection schemes. The situation of Ireland is analyzed in \cite{2_10k, 2_10l, 2_10m}; in particular, authors in \cite{2_10k} compare three mitigation strategies - RoCoF relay setting relaxation, minimum inertia level in the market and provision of synthetic inertia - with dynamic simulations, conveying that relaxing the settings of RoCoF relays has the most influence in minimizing the risk of frequency instability. Similar studies have been carried out also for the UK, showing that with the current requirements for FCR the frequency nadir and RoCoF exceed respectively 49.0 Hz and 0.125 Hz/s \cite{2_10n, 2_10p}.

Besides conventional generators, frequency support can be provided by other components capable of injecting active power into the grid, e.g. High-Voltage Direct Current (HVDC) lines. According to \cite{2_10}, the control scheme of all HVDC converters must be capable of operating in frequency sensitive mode, i.e. the transmitted power is adjusted in response to a frequency deviation. For this reason, an HVDC link connecting asynchronous areas can be used as a vehicle for FCR-D: to limit the IFD in case of a disturbance, the necessary active power can be imported from the neighboring system in the form of Emergency Power Control (EPC). Given the high number of interconnections formed by HVDC lines between NSA and the neighboring areas (see \figurename~\ref{fig:1_map}) and the introduction in the Nordic market of a new FFR product expected by summer 2020, this \textit{corrective} action could represent a valid alternative to expensive \textit{preventive} redispatching. The current EPC activation method is based on step-wise triggers: when the frequency drops below a certain threshold, a constant amount of power is injected through the HVDC link, depending on the level of inertia. Although already implemented \cite{2_9}, HVDC EPC is currently not in use. %
% \begin{figure}[!t]
%     \centering
%     % \includegraphics[trim = 0.3cm 14.7cm 9.2cm 0.2cm,clip,width=0.46\textwidth]{Figures/Files/Map.pdf}
%     \includegraphics[width=0.46\textwidth]{Figures/Files/Map2.pdf}
%     %\vspace{-1em}
%     \caption{Regional Groups, HVDC interconnections and dimensioning incident (Oskarshamn 3) \cite{2_11}, \cite{2_12}.\label{fig:1_map}}
% %\vspace{-1.0em}
% \end{figure}

\definecolor{blueMAP}{rgb}{0.1098,0.2039,0.4235}%
\definecolor{orangeMAP}{rgb}{0.9608,0.6157,0.3373}%
\definecolor{purpleMAP}{rgb}{0.3294,0.2588,0.4157}%

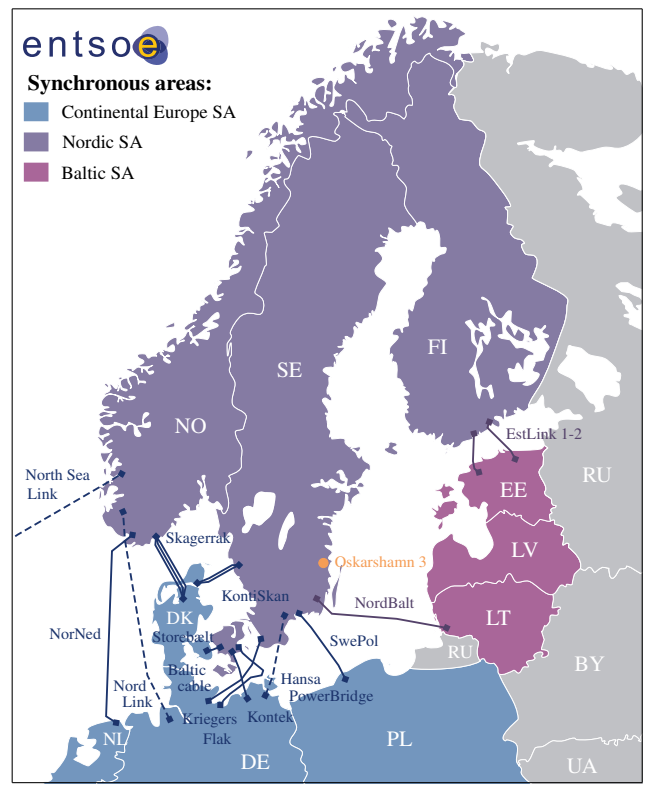
\begin{figure}[!t]
    \centering
    \begin{tikzpicture}
        \node[inner sep=0pt, anchor = south west] (network) at (0,0) {\includegraphics[trim = 0.6cm 0.6cm 0.6cm 0.5cm,clip,width=0.46\textwidth]{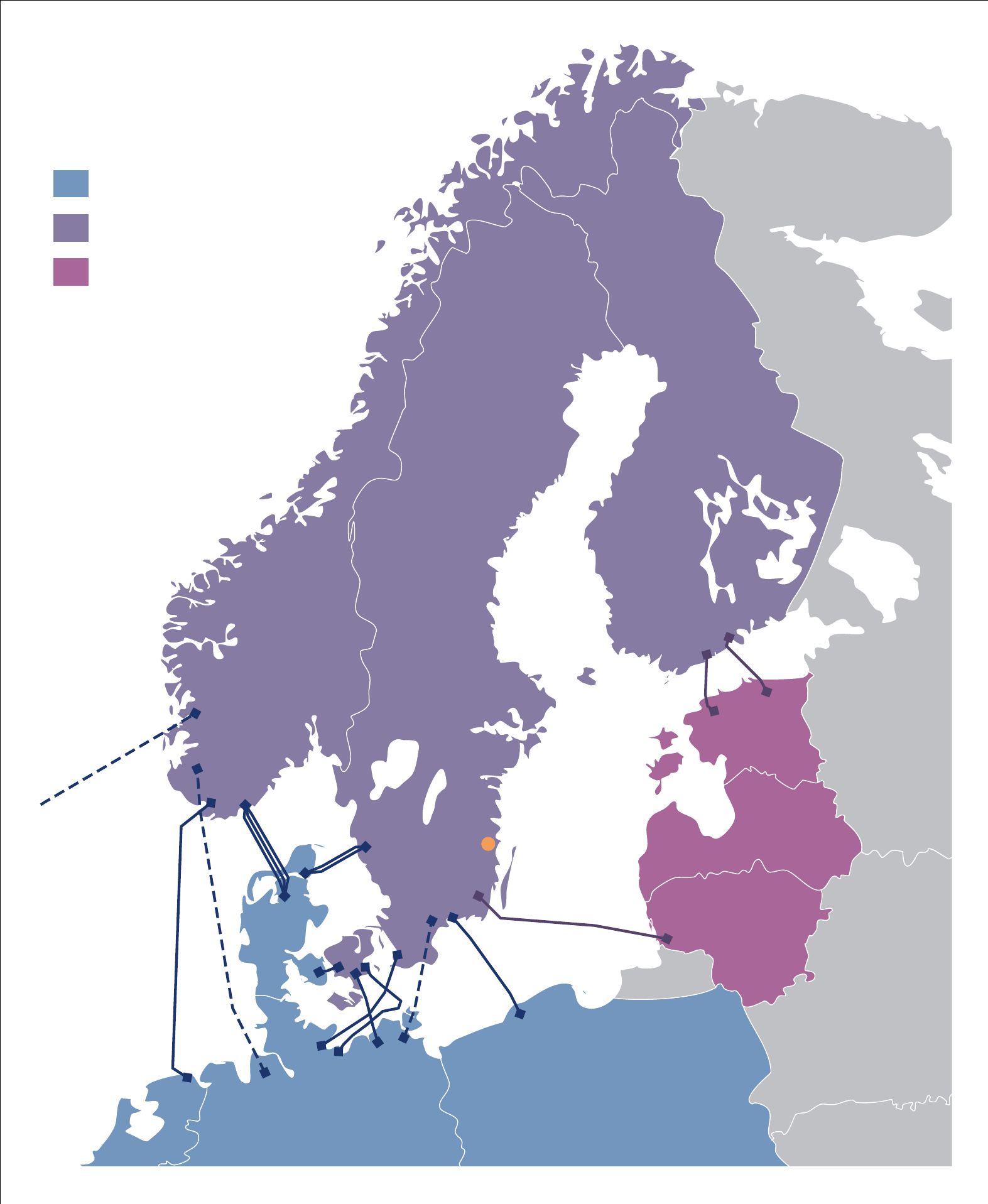}};
        \node[inner sep=0pt, anchor = south west] (network) at (0.1,9.5) {\includegraphics[trim = 0.35cm 0.4cm 0.4cm 0.4cm,clip,width=0.11\textwidth]{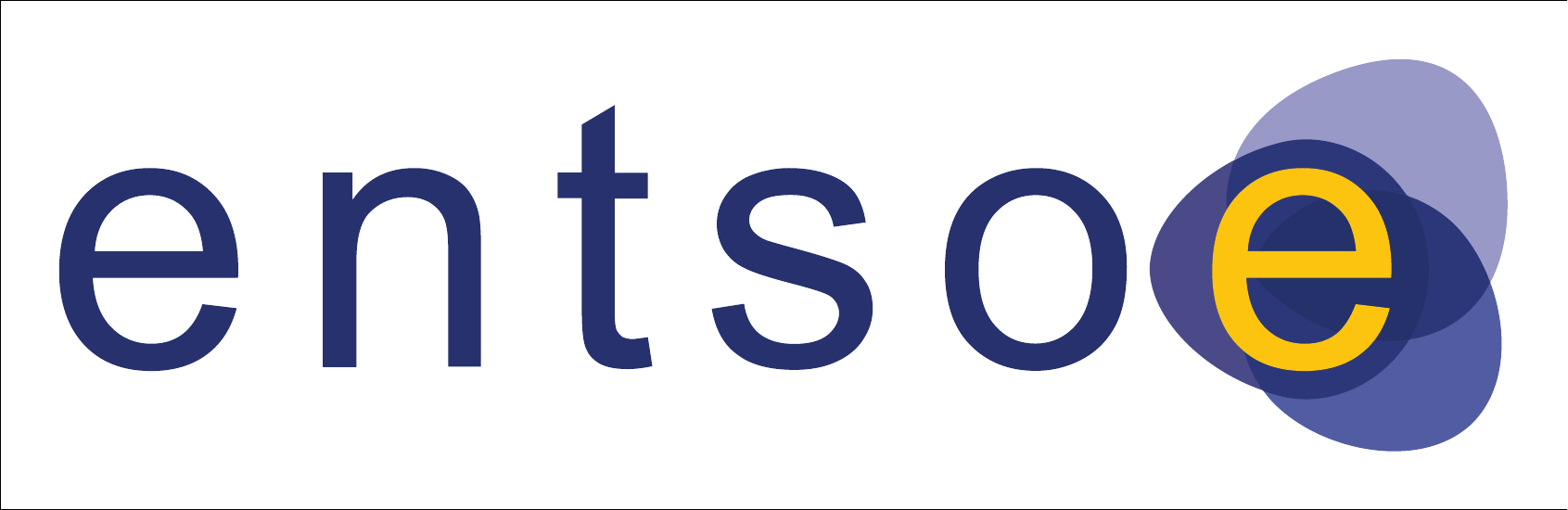}};
        \draw (0,0.02) -- (8.29,0.02);
        \draw (8.29,0.02) -- (8.29,10.28);
        \draw (8.29,10.28) -- (0,10.28);
        \draw (0,0.02) -- (0,10.28);
        \node[] at (1.42,9.28) {\footnotesize  \textbf{Synchronous areas:}};
        \node[] at (1.80,8.89) {\scriptsize  Continental Europe SA};
        \node[] at (1.19,8.515) {\scriptsize  Nordic SA};
        \node[] at (1.13,8.1) {\scriptsize  Baltic SA};
        \node[] at (1.36,0.6) {\scriptsize   \textcolor{white}{NL}};
        \node[] at (3.65,5.5) {\footnotesize   \textcolor{white}{SE}};
        \node[] at (3.2,0.3) {\footnotesize   \textcolor{white}{DE}};
        \node[] at (5.1,0.6) {\footnotesize   \textcolor{white}{PL}};
        \node[] at (6.4,2.2) {\footnotesize   \textcolor{white}{LT}};
        \node[] at (6.75,3.1) {\footnotesize   \textcolor{white}{LV}};
        \node[] at (6.6,3.9) {\footnotesize   \textcolor{white}{EE}};
        \node[] at (7.7,4.1) {\footnotesize   \textcolor{white}{RU}};
        \node[] at (5.9,1.75) {\scriptsize   \textcolor{white}{RU}};
        \node[] at (2.21,2.2) {\scriptsize  \textcolor{white}{DK}};
        \node[] at (5.6,5.8) {\footnotesize  \textcolor{white}{FI}};
        \node[] at (2.35,4.75) {\footnotesize  \textcolor{white}{NO}};
        \node[] at (7.6,1.6) {\footnotesize   \textcolor{white}{BY}};
        \node[] at (7.5,0.25) {\footnotesize   \textcolor{white}{UA}};
        \node[] at (0.83,2) {\tiny  \textcolor{blueMAP}{NorNed}};
        \node[] at (0.6,4.1) {\tiny  \textcolor{blueMAP}{North Sea}};
        \node[] at (0.4,3.85) {\tiny  \textcolor{blueMAP}{Link}};
        \node[] at (2.45,3.25) {\tiny  \textcolor{blueMAP}{Skagerrak}};  
        \node[] at (3.2,2.5) {\tiny  \textcolor{blueMAP}{KontiSkan}};
        \node[] at (2.25,1.96) {\tiny  \textcolor{blueMAP}{Storeb{\ae}lt}};
        \node[] at (2.3,1.52) {\tiny  \textcolor{blueMAP}{Baltic}};
        \node[] at (2.4,1.3) {\tiny  \textcolor{blueMAP}{cable}};
        \node[] at (1.55,1.3) {\tiny  \textcolor{blueMAP}{Nord}};
        \node[] at (1.65,1.1) {\tiny  \textcolor{blueMAP}{Link}};
        \node[] at (2.6,0.85) {\tiny  \textcolor{blueMAP}{Kriegers}};
        \node[] at (2.7,0.6) {\tiny  \textcolor{blueMAP}{Flak}};
        \node[] at (3.8,1.4) {\tiny  \textcolor{blueMAP}{Hansa}};
        \node[] at (4.2,1.15) {\tiny  \textcolor{blueMAP}{PowerBridge}};
        \node[] at (3.4,0.9) {\tiny  \textcolor{blueMAP}{Kontek}};
        \node[] at (4.5,1.9) {\tiny  \textcolor{blueMAP}{SwePol}};
        \node[] at (4.85,2.95) {\tiny  \textcolor{orangeMAP}{Oskarshamn 3}};
        \node[] at (4.9,2.4) {\tiny  \textcolor{purpleMAP}{NordBalt}};
        \node[] at (7,4.65) {\tiny  \textcolor{purpleMAP}{EstLink 1-2}};
    \end{tikzpicture}
    \caption{Synchronous areas, HVDC interconnections and dimensioning incident (Oskarshamn 3) \cite{2_11}, \cite{2_12}. Dashed lines are expected to be operational by 2025.}
\label{fig:1_map}
\vspace{-0.8em}
\end{figure}%

The utilization of HVDC interconnectors for frequency support has been largely investigated from a technical point of view \cite{2_13, 2_14, 2_15, 2_16, 2_17, 2_18}; however, limited work has been done on the evaluation of the related economic benefits. This is the case also for other mitigation strategies, e.g. \cite{2_10j,2_10k, 2_10l, 2_10m}, which are in general analyzed based on their technical properties. The only document referring to the potential costs of these remedial actions is \cite{2_9}; however the discussion always remains qualitative. In this regard, the contributions of this paper are:

\addvbuffer[4pt 6pt]{%
\setlength{\extrarowheight}{2pt}
\hskip-1.5em
\begin{tabularx}{0.488\textwidth}{l *1{>{\arraybackslash}X}}
\vspace{-0.1em}
     $\bullet$ & a quantitative analysis on the costs of different remedial actions, focusing on the EPC functionality of HVDC links for the provision of frequency support; \\
     $\bullet$ & the design of a decision-making support tool for deciding how to act in case of low inertia situations, not only from a technical but also from an economic point of view; \\
     $\bullet$ & the comparison of different mitigation strategies and the corresponding cost savings on the Nordic Synchronous Area. \\
\end{tabularx}%
}

The analysis is carried out for three scenarios (2018, 2020, 2025), using historical data from Nord Pool (2018), and inertia estimates from the Nordic TSOs (2020-2025). The costs of the remedial actions are calculated based on historical data from the past 5 years, considering the distribution of the average price obtained via data re-sampling, making this paper the first effort to calculate these costs in a quantitative manner.

The rest of the paper is organized as follows. \mbox{Section \ref{sec:2}} provides a brief overview of the Nordic power system. \mbox{Section \ref{sec:3}} introduces the decision-making support tool and which data is used for the three different scenarios. \mbox{Section \ref{sec:3_0}} presents the under frequency regression model used to determine the IFD based on the kinetic energy of the system. \mbox{Section \ref{sec:4}} describes in detail the current paradigm and some alternative remedial actions, focusing on some technical considerations. \mbox{Section \ref{sec:5}} introduces the market considerations and price scenarios used for the calculation of the related costs. \mbox{Section \ref{sec:6}} presents the cost saving analyses and \mbox{Section \ref{sec:7}} concludes.

\section{Nordic Synchronous Area}\label{sec:2}
A synchronous area (SA), also referred to as regional group, is a group of power systems that are connected and operate under the same frequency. In Europe, there are five main synchronous areas \cite{2_12}:

\addvbuffer[4pt 6pt]{%
\setlength{\extrarowheight}{2pt}
\hskip-1.5em
\begin{tabularx}{0.488\textwidth}{l *1{>{\arraybackslash}X}}
\vspace{-0.1em}
     $\bullet$ & Continental Europe SA; \\
     $\bullet$ & Nordic SA; \\
     $\bullet$ & Baltic SA; \\
     $\bullet$ & British SA; \\
     $\bullet$ & Ireland/Northern Ireland SA.
\end{tabularx}%
}

The system of interest – Nordic SA, or NSA – consists of four countries: Norway, Sweden, Finland and East Denmark (see \figurename~\ref{fig:1_map}). West Denmark is synchronously connected to Germany, thus belongs to Continental Europe SA. The generation mix in the Nordic countries can be found in the \mbox{ENTSO-E} Transparency Platform \cite{2_20}. Almost half of the generation in East Denmark comes from renewable sources (onshore and offshore wind, solar, biomass and waste), while the remaining is fossil fuel based (natural gas, coal and oil). In Norway, more than 90\% of electricity is produced by hydro power plants. Hydro power plants contribute to half of the generation in Sweden as well, the remaining capacity is divided between nuclear power plants, wind farms and oil-fired thermal units. In Finland the generation mix is more heterogeneous; half of the Finnish electrical energy is produced by nuclear power plants and coal-based thermal units.

The grand total installed generation capacity was 91.66 GW in 2018, and it has gradually increased to 97.2 GW in the last two years. Overall, the region is water resource rich, hence majority (about 50\%) of electricity production comes from hydro power plants. The second source of generation is thermal (about 20\%) and nuclear comes third (about 12\%). The penetration of RES has gradually increased from 11.1 GW (12\%) in 2018 to 14.5 GW (15\%) in 2020, as Sweden and Norway are installing a large number of onshore wind turbines. The total demand ranges from 26.4 to 60.1~GW, with the minimum and the maximum respectively during summer and winter. With a similar trend, the total system kinetic energy ranges from 100 to 300~GWs. Based on the system data, NSA is comparable to the British SA and it is not immune to high frequency deviations during low-inertia operation.

In terms of interconnections to neighboring asynchronous areas, NSA is connected to Continental Europe SA through seven HVDC links: NorNed (Norway-Netherlands, 700 MW), Skagerrak (Norway-West Denmark, 1'632 MW), KontiSkan (Sweden-West Denmark, 720~MW), Storeb{\ae}lt (West Denmark-East Denmark, 600 MW), Kontek (East Denmark-Germany, 600 MW), Baltic cable (Sweden-Germany, 600 MW) and SwePol (Sweden-Poland, 600 MW). In addition, Kriegers Flak (Combined Grid Solution) will provide connection between East Denmark and Germany starting from August 2020 (AC cable and back-to-back HVDC converter, 400 MW), integrating offshore wind farms along its path. Finally, three other HVDC links connect NSA to Baltic SA: NordBalt (Sweden-Lithuania), Estlink (Finland-Estonia) and Vyborg HVDC (Finland-Russia).

\subsection{Future Scenarios}\label{ssec:2_1}
As mentioned in several Nordic TSOs reports \cite{2_9, 3_1}, the Nordic power system is undergoing substantial changes. The main drivers are climate policies, which in turn stimulate the development of more Renewable Energy Sources (RES), technological developments, and a common European framework for markets, operation and planning. As a consequence, the following events are expected in the next years \cite{3_1}:

\addvbuffer[4pt 6pt]{%
\setlength{\extrarowheight}{2pt}
\hskip-1.5em
\begin{tabularx}{0.488\textwidth}{l *1{>{\arraybackslash}X}}
\vspace{-0.1em}
     $\bullet$ & Closure of fossil-fuelled thermal power plants and reduction of CHP plants; \\
     $\bullet$ & Installation of a large amount of wind power capacity to meet political agreements; \\
     $\bullet$ & Decommissioning of Swedish nuclear power plants and construction of new nuclear power plants in Finland; \\
     $\bullet$ & Construction of new HVDC interconnectors between NSA and neighboring areas.
\end{tabularx}%
}
In order to assess the impact of the ongoing and foreseen changes on the kinetic energy of the system, future market scenarios have been defined by the Nordic TSOs. In \cite{2_9}, a scenario for 2025 is presented: the scenario is constructed as a best-estimate scenario, i.e. it describes the market development that is considered most probable. 

In brief, the grand total installed generation capacity is expected to increase by 6.2 GW. The new capacity from RES is expected to be 8.44 GW, of which 6.72 GW of wind power, while the closure of fossil-fuelled thermal units will decrease synchronous generation by 2.24 GW. The demand is expected to increase by 6\%, mainly due to growing population and electrification of vehicles and railroads. 

In terms of new connections, three HVDC links are expected to become operational between 2020 and 2025 (dashed lines in \figurename~\ref{fig:1_map}): North Sea Link (Norway-UK, 1'400 MW) and Nord Link (Norway-Germany, 1'400 MW) in 2021 and Hansa PowerBridge (Sweden-Germany, 700 MW) in 2025 \cite{3_1}. 
\section{Decision-Making Support Tool}\label{sec:3}

\definecolor{blueCHART1}{rgb}{0.1216,0.3059,0.4745}%
\definecolor{blueCHART2}{rgb}{0.3569,0.6078,0.8353}%
\definecolor{orangeCHART1}{rgb}{0.9294,0.4902,0.1922}%
\definecolor{orangeCHART2}{rgb}{1.0000,0.7529,0}%

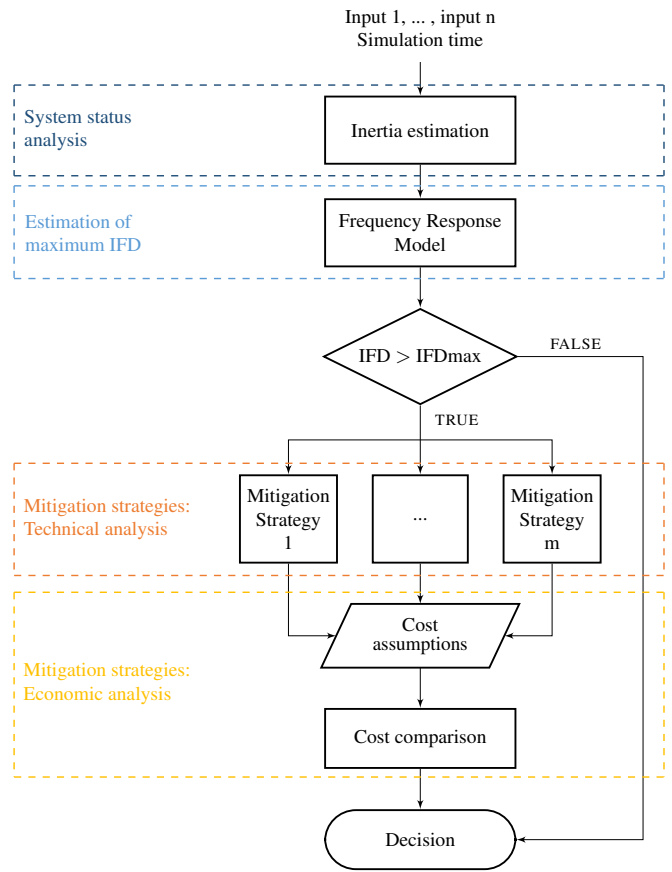
\begin{figure}
\centering
\resizebox{0.48\textwidth}{!}{%
\begin{tikzpicture}[%
    >=latex',
    every node/.append style={font=\footnotesize}, 
    node distance=5mm,
    datashape/.style={
    trapezium, draw, 
    trapezium left angle=65,
    trapezium right angle=-65,
    outer sep=0pt,
    minimum height = 27pt,
    text width=40pt}
    ]
    % help grid
    % \draw[help lines,step=.2] (-6,-13) grid (4,1);
    % \draw[help lines,line width=.6pt,step=1] (-6,-13) grid (4,1);
    
    % BLOCKS
    \node [block, draw=none, execute at begin node=\setlength{\baselineskip}{10pt}] (origin) {Input 1, ... , input n \\ Simulation time};
    \node [block, line width = 0.8pt, minimum width = 80pt, below = of origin] (estim) {Inertia estimation};
    \node [block, line width = 0.8pt, minimum width = 80pt, below = of estim, execute at begin node=\setlength{\baselineskip}{10pt}] (model) {Frequency Response \\ Model};
    \node [draw, diamond, line width = 0.8pt, below= 6 mm of model,  aspect=2] (if) {IFD $>$ IFDmax};
    \node [below= 4 mm of if] (inter1) {};  
    \node [block, minimum height = 37pt, line width = 0.8pt, below= 4 mm of inter1] (mit) {...};
    \node [block, minimum height = 37pt, line width = 0.8pt, left= 5 mm of mit, execute at begin node=\setlength{\baselineskip}{10pt}] (mit_1) {Mitigation \\ Strategy \\ 1};
    \node [block, minimum height = 37pt, line width = 0.8pt, right= 5 mm of mit, execute at begin node=\setlength{\baselineskip}{10pt}] (mit_n) {Mitigation \\ Strategy \\ m};
    \node [right= 5 mm of mit_n] (inter2) {};  
    \node [datashape, execute at begin node=\setlength{\baselineskip}{8pt}, line width = 0.8pt, below = 6 mm of mit] (assump) {\centering Cost \\ assumptions};
    \node [block, minimum height = 25pt, minimum width = 80pt, line width = 0.8pt, below= 6 mm of assump] (comp) {Cost comparison};
    \node [block, minimum height = 25pt, minimum width = 80pt, rounded corners=13pt, line width = 0.8pt, below= 6 mm of comp] (decision) {Decision};
    
    \node [block, draw=none, left =27.3mm of estim, execute at begin node=\setlength{\baselineskip}{10pt}, align=left] (step1) {\color{blueCHART1} System status \\ \color{blueCHART1} analysis};
    \node [block, draw=none, above right = 6mm and 8mm of step1.center] (step1_1) {};
    \node [block, draw=none, below right = 6mm and 8mm of step1.center] (step1_2) {};
    \node [block, draw=none, left =25.9mm of model, execute at begin node=\setlength{\baselineskip}{10pt}, align=left] (step2) {\color{blueCHART2} Estimation of \\ \color{blueCHART2} maximum IFD};
    \node [block, draw=none, above right = 6mm and 7.2mm of step2.center] (step2_1) {};
    \node [block, draw=none, below right = 6mm and 7.2mm of step2.center] (step2_2) {};
    \node [block, draw=none, left =6mm of mit_1, execute at begin node=\setlength{\baselineskip}{10pt}, align=left] (step3) {\color{orangeCHART1} Mitigation strategies: \\ \color{orangeCHART1} Technical analysis};
    \node [block, draw=none, above right =8mm and 9.7mm of step3.center] (step3_1) {};
    \node [block, draw=none, below right =8mm and 9.7mm of step3.center] (step3_2) {};
    \node [block, draw=none, below left =2mm and 32.8mm of assump.center, execute at begin node=\setlength{\baselineskip}{10pt}, align=left] (step4) {\color{orangeCHART2} Mitigation strategies: \\ \color{orangeCHART2} Economic analysis};
    \node [block, draw=none, above right =14mm and 9mm of step4.center] (step4_1) {};
    \node [block, draw=none, below right =15mm and 9mm of step4.center] (step4_2) {};
    
    %CONNECTORS
    \draw [->] (origin)--(estim);
    \draw [->] (estim)--(model);
    \draw [->] (model)--(if);
    \draw [->] (if)--node[above right = 1mm and 1mm]{\textsc{true}}(mit);
    \draw [->] (inter1.center)-|(mit_1);
    \draw [->] (inter1.center)-|(mit_n);
    \draw [->] (mit)--(assump);
    \draw [->] (mit_1)|-(assump);
    \draw [->] (mit_n)|-(assump);
    \draw [->] (assump)--(comp);
    \draw [->] (comp)--(decision);
    \draw [->] (if) -| node[above left = 0mm and 5mm]{\textsc{false}} (inter2.center)|-(decision);
    \draw [blueCHART1, dashed, line width=.6pt] (-6,-0.85)--(3.6,-0.85);
    \draw [blueCHART1, dashed, line width=.6pt] (-6,-0.85)--(-6,-2.2);
    \draw [blueCHART1, dashed, line width=.6pt] (3.6,-0.85)--(3.6,-2.2);
    \draw [blueCHART1, dashed, line width=.6pt] (-6,-2.2)--(3.6,-2.2);
    \draw [blueCHART2, dashed, line width=.6pt] (-6,-2.35)--(3.6,-2.35);
    \draw [blueCHART2, dashed, line width=.6pt] (-6,-2.35)--(-6,-3.73);
    \draw [blueCHART2, dashed, line width=.6pt] (3.6,-2.35)--(3.6,-3.73);
    \draw [blueCHART2, dashed, line width=.6pt] (-6,-3.73)--(3.6,-3.73);
    \draw [orangeCHART1, dashed, line width=.6pt] (-6,-6.48)--(3.6,-6.48);
    \draw [orangeCHART1, dashed, line width=.6pt] (-6,-6.48)--(-6,-8.15);
    \draw [orangeCHART1, dashed, line width=.6pt] (-6,-8.15)--(3.6,-8.15);
    \draw [orangeCHART1, dashed, line width=.6pt] (3.6,-6.48)--(3.6,-8.15);
    \draw [orangeCHART2, dashed, line width=.6pt] (-6,-8.4)--(3.6,-8.4);
    \draw [orangeCHART2, dashed, line width=.6pt] (-6,-8.4)--(-6,-11.15);
    \draw [orangeCHART2, dashed, line width=.6pt] (3.6,-8.4)--(3.6,-11.15);
    \draw [orangeCHART2, dashed, line width=.6pt] (-6,-11.15)--(3.6,-11.15);
    % \draw [decorate,decoration={brace,amplitude=3pt,mirror},line width = 0.8pt,blueCHART1] (step1_1) -- (step1_2);
    % \draw [decorate,decoration={brace,amplitude=3pt,mirror},line width = 0.8pt,blueCHART2] (step2_1) -- (step2_2);
    % \draw [decorate,decoration={brace,amplitude=3pt,mirror},line width = 0.8pt,orangeCHART1] (step3_1) -- (step3_2);
    % \draw [decorate,decoration={brace,amplitude=3pt,mirror},line width = 0.8pt,orangeCHART2] (step4_1) -- (step4_2);
\end{tikzpicture}
}%
\caption{Sequence diagram of the algorithm comparing different mitigation strategies.}
\vspace{-1.0em}
\label{fig:2_diagram}
\end{figure}

The ongoing decrease of system inertia is classified as one of the major future challenges for the Nordic Power System \cite{3_1}. For this reason, all Nordic TSOs (Energinet - Denmark, Svenska kraftn{\"a}t - Sweden, Statnett - Norway, and Fingrid - Finland) have implemented in their SCADA systems an online tool for estimating the inertia of the system and the corresponding maximum IFD \cite{2_9}. These tools are used to estimate the inertia level, and the corresponding maximum IFD, from seven to one day before operation based on production plans; however, they can be used also for forecasting future inertia levels based on market simulations. 

In this paper, we propose an algorithm which extends these online-estimation tools. First, the inertia level of the system and the corresponding IFD are estimated. If the frequency limits and N-1 security criterion are violated, different mitigation strategies are then compared, both from a technical and an economic point of view. The flow chart of the proposed method is depicted in \figurename~\ref{fig:2_diagram}. 

The first block relates to the estimation of system inertia. This calculation is performed considering which units are online, or will be online. There could be different input for this, such as hourly production values from power exchanges, production measurements per production type, measurements of power plants, status of generators' circuit breaker, and so on. The length of the simulation period and the time resolution are also defined at this stage, meaning that the analysis can be performed for the next hour, but also for longer periods (e.g. an entire year) with hourly, or longer, time resolution.

The second block corresponds to the estimation of the maximum IFD. For this calculation, the frequency response model of the system under investigation is used. This can be a dynamic, e.g. single- or multi-machine equivalent, or a statistical model. The input to this block is the frequency of the system, the reference incident and the inertia estimates (the kinetic energy reduction due to the reference incident can be considered as well). 

Once the maximum IFD is calculated with the desired time resolution, these values are compared to the maximum allowed IFD of the system. If the requirements are not violated, then there is no need for further analyses and the algorithm goes directly to the conclusion; otherwise, different mitigation strategies are compared in the following blocks.

A large variety of mitigation strategies can be applied in order to guarantee N-1 security in low-inertia situations \cite{2_9, 2_10j}. Depending on the inertia level of the system, the entity of these actions could vary significantly. The scope of the technical analysis is to calculate the entity of each action, e.g. how much the reference incident should be reduced, or how much extra support to FCR-D is necessary. This calculation is normally performed and validated with dynamic simulations. 

After the technical analysis has been performed, the cost of each strategy is calculated taking into consideration the different market stages where these actions are performed. The input to this blocks are different market considerations and the expected evolution of prices in the upcoming period. Once the costs are calculated, it is possible to compare the analyzed strategies from an economic point of view and choose the one which results in least costs. 

In this paper, the cost saving analysis starts with the events of 2018, and continues with two scenarios for the years 2020 and 2025. As power system data is considered sensitive by Nordic TSOs, the comparison of the two remedial actions considered in this paper varies across scenarios based on the availability of data. The three scenarios and the corresponding data are further described in this section.

\vspace{-0.5em}
\subsection{Summer 2018}\label{ssec:3_1}
During Summer 2018, the inertia of the Nordic System dropped below the safety level three times. The length of the periods and the limitations on the largest unit, O3, have been communicated by Svk through Urgent Market Messages (UMM) in the Nord Pool Online Platform \cite{2_8}. The three periods are:

\addvbuffer[4pt 6pt]{%
\setlength{\extrarowheight}{2pt}
\hskip-1.5em
\begin{tabularx}{0.488\textwidth}{l *1{>{\arraybackslash}X}}
\vspace{-0.1em}
     -- & June 23-25: duration 50 hours, dimensioning incident reduced by 100 MW; \\
     -- & July 6-9: duration 75 hours, dimensioning incident reduced by 100 MW; \\
     -- & August 11-12: duration 41 hours, dimensioning incident reduced by 100 MW.
\end{tabularx}%
}

During these periods, the kinetic energy of the system was not always below the security threshold; however, for security reasons and technical limitations (ramping limits and costs), the output of O3 was reduced for the entire length of these periods. As the duration of these events and the entities of the reduction are known, only the comparison of the two remedial action is performed for this scenario (thus starting directly from the technical analysis).

\subsection{Scenarios 2020 and 2025}\label{ssec:3_2}
The kinetic energy stored in the system, $E_k$, is proportional to the rated apparent power of synchronous generators connected to the grid, $S$, and their inertia constants, $H$:
\begin{equation}\label{eq:3_2}
    E_k = \sum_{i=1}^N H_i S_i
\end{equation}
Thus, in order to estimate the kinetic energy in the system, it is necessary to know which generating units will be online. To this end, Svk has performed a large number of market simulations for the upcoming years \cite{3_4, 3_5}. The resulting market outcomes, i.e. the production of generating units and the flows over HVDC interconnectors, have been used to estimate the inertia level of the system for these years.

The market simulations are carried out using the EFI's Multi-area Power-market Simulator (EMPS) with the SAMLAST functionality \cite{2_9}. The EMPS model is a market simulation tool, designed for power systems with high shares of hydro power plants, which comprises a strategy evaluation and a system simulation part \cite{3_1a}. The strategy evaluation consists in the calculation of incremental water values (marginal costs for hydro power) for each area using stochastic dynamic programming. These values are then used in the simulation part, where an optimization problem is solved to determine the production of hydro and thermal power plants. The generators dispatch is then analyzed with the SAMLAST program, which performs load-flow analyses using detailed grid models \cite{3_1b}. If grid constraints are violated, SAMLAST reduces the available transmission capacities between areas and update the optimization problem with the new capacities. In principle, the final solution will coincide with the outcome of a well-functioning electricity market. 

The market model used for the simulation comprises more than 1'200 hydro power plants and all thermal units with capacity greater than 100 MW \cite{2_9}. The countries included in the model are Sweden, Denmark, Norway, Finland, Latvia, Lithuania and Estonia, while the neighboring countries (Germany, Poland, the Netherlands and the UK) are described with fixed prices \cite{2_9}. For the 2020 scenario, revisions of nuclear power plants is taken into consideration \cite{3_1c, 3_1d, 3_1e}; since these events are scheduled 2 years in advance, they are not accounted for in the 2025 scenario. Instead, two different scenarios are considered for the year 2025 based on the decommission of nuclear power plants:

\addvbuffer[4pt 6pt]{%
\setlength{\extrarowheight}{4pt}
\hskip-1.8em
\begin{tabular}{m{1pt} p{0.455\textwidth}}
     1. & \textit{2025 - Full nuclear (FN)}: market simulations based on the current situation, with nuclear power plants fully dispatched. \\
     2. & \textit{2025 - Half nuclear (HN)}: market simulations with half of the nuclear production replaced by wind and solar production and HVDC imports (all inverter-based generation).
\end{tabular}%
}
\noindent
These scenarios are further described in a previous report from ENTSO-E \cite{3_3}, and comprise the foreseen changes introduced in Section \ref{ssec:2_1}, i.e. estimations on future wind power installed capacity, information on synchronous generation plants decommission and new HVDC interconnections.

\definecolor{blueCHART1}{rgb}{0.1216,0.3059,0.4745}%
\definecolor{blueCHART2}{rgb}{0.3569,0.6078,0.8353}%
\definecolor{orangeCHART1}{rgb}{0.9294,0.4902,0.1922}%
\definecolor{orangeCHART2}{rgb}{1.0000,0.7529,0}%

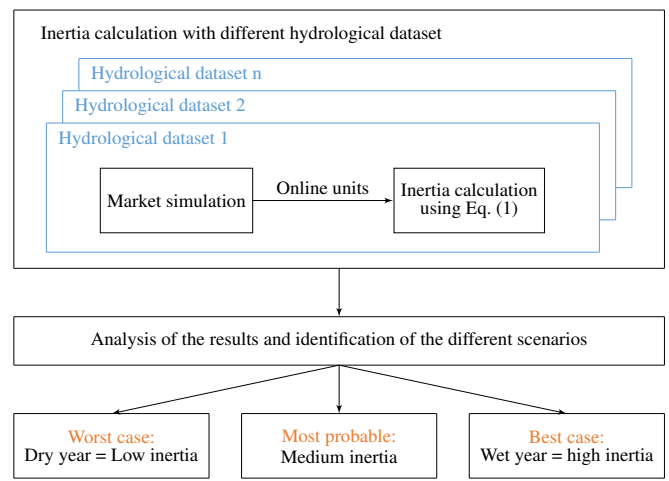
\begin{figure}[!b]
\centering
\resizebox{0.48\textwidth}{!}{%
\begin{tikzpicture}[%
    >=latex',
    every node/.append style={font=\footnotesize}, 
    node distance=5mm,
    datashape/.style={
    trapezium, draw, 
    trapezium left angle=65,
    trapezium right angle=-65,
    outer sep=0pt,
    minimum height = 27pt,
    text width=40pt}
    ]
    % help grid
    % \draw[help lines,step=.2] (-6,-7) grid (4,1);
    % \draw[help lines,line width=.6pt,step=1] (-6,-7) grid (4,1);
    
    % BLOCKS
    \node [draw=none] at (-2.5,0.375) {Inertia calculation with different hydrological dataset};
    
    \node [draw=none] at (-4,-1.25) {\textcolor{blueCHART2}{Hydrological dataset 1}};
    \node [draw=none] at (-3.75,-0.75) {\textcolor{blueCHART2}{Hydrological dataset 2}};
    \node [draw=none] at (-3.50,-0.25) {\textcolor{blueCHART2}{Hydrological dataset n}};
    
    \node [block] at (-3.5,-2.2) {Market simulation};
    \node [draw=none] at (-1.25,-2) {Online units};
    \node [block, execute at begin node=\setlength{\baselineskip}{8pt}] at (1,-2.2) {Inertia calculation \\ using Eq. (1)};
    
    \node [draw=none] at (-1,-4.375) {Analysis of the results and identification of the different scenarios};
    
    \node [draw=none] at (-4.5,-5.87) {\textcolor{orangeCHART1}{Worst case:}};
    \node [draw=none, execute at begin node=\setlength{\baselineskip}{8pt}] at (-4.5,-6.17) {Dry year = Low inertia};
    
    \node [draw=none] at (-1,-5.88) {\textcolor{orangeCHART1}{Most probable:}};
    \node [draw=none, execute at begin node=\setlength{\baselineskip}{8pt}] at (-1,-6.17) {Medium inertia};
    
    \node [draw=none] at (2.5,-5.87) {\textcolor{orangeCHART1}{Best case:}};
    \node [draw=none, execute at begin node=\setlength{\baselineskip}{8pt}] at (2.5,-6.17) {Wet year = high inertia};
    
    %CONNECTORS
    \draw [-] (-6,0.75)--(4,0.75);
    \draw [-] (-6,-3.25)--(4,-3.25);
    \draw [-] (-6,0.75)--(-6,-3.25);
    \draw [-] (4,0.75)--(4,-3.25);
    
    \draw [blueCHART2] (-5.5,-1)--(3,-1);
    \draw [blueCHART2] (-5.5,-3)--(3,-3);
    \draw [blueCHART2] (-5.5,-1)--(-5.5,-3);
    \draw [blueCHART2] (3,-1)--(3,-3);
    
    \draw [blueCHART2] (-5.25,-0.5)--(3.25,-0.5);
    \draw [blueCHART2] (-5.25,-0.5)--(-5.25,-1);
    \draw [blueCHART2] (3.25,-0.5)--(3.25,-2.5);
    \draw [blueCHART2] (3,-2.5)--(3.25,-2.5);
    
    \draw [blueCHART2] (-5,0)--(3.50,0);
    \draw [blueCHART2] (-5,0)--(-5,-0.5);
    \draw [blueCHART2] (3.5,0)--(3.5,-2);
    \draw [blueCHART2] (3.25,-2)--(3.5,-2);
    
    \draw [->] (-2.33,-2.2)--(-0.17,-2.2);
    
    \draw [->] (-1,-3.25)--(-1,-4);
    
    \draw [-] (-6,-4)--(4,-4);
    \draw [-] (-6,-4.75)--(4,-4.75);
    \draw [-] (-6,-4)--(-6,-4.75);
    \draw [-] (4,-4)--(4,-4.75);
    
    \draw [->] (-1,-4.75)--(-1,-5.5);
    \draw [->] (-1,-4.75)--(-4.5,-5.5);
    \draw [->] (-1,-4.75)--(2.5,-5.5);
    
    \draw [-] (-6,-5.5)--(-3,-5.5);
    \draw [-] (-6,-6.5)--(-3,-6.5);
    \draw [-] (-6,-5.5)--(-6,-6.5);
    \draw [-] (-3,-5.5)--(-3,-6.5);
    
    \draw [-] (-2.5,-5.5)--(0.5,-5.5);
    \draw [-] (-2.5,-6.5)--(0.5,-6.5);
    \draw [-] (-2.5,-5.5)--(-2.5,-6.5);
    \draw [-] (0.5,-5.5)--(0.5,-6.5);
    
    \draw [-] (1,-5.5)--(4,-5.5);
    \draw [-] (1,-6.5)--(4,-6.5);
    \draw [-] (1,-5.5)--(1,-6.5);
    \draw [-] (4,-5.5)--(4,-6.5);
\end{tikzpicture}
}%
\caption{Determination of inertia scenarios based on simulations with different meteorological data.}
\vspace{-1.0em}
\label{fig:3_inertiadiagram}
\end{figure}

As there is a strong correlation between weather conditions and system inertia, e.g. in dry periods with low rainfall the production of hydro power plants is replaced by HVDC imports leading to low inertia levels \cite{3_1}, the weather conditions assumed for 2020 and 2025 highly impact the results. In order to capture a greater variety of possible conditions, each market simulation (for the years 2020 and 2025) is performed several times; each time, different meteorological conditions are applied (see \figurename~\ref{fig:3_inertiadiagram}). The meteorological dataset used by Svk comprises the years between 1980 and 2012. Each simulation is run for a time-series corresponding to the whole year with a resolution of 3 hours \cite{2_9}. 

The validation of the market model was performed comparing the outcome of the simulation for the year 2014 with the actual data \cite{2_9}. In general, the model captures the seasonal behaviour of generation of both hydro and thermal units, with some minor errors. For example, the production of thermal plants is underestimated during summer because the model does not comprise start-up costs and reserve requirements, which results in thermal units to stop producing for short periods.

The outcome of the market simulations determines which units are online, which in turn are used to estimate the kinetic energy in the system at each time instance. For each time instance, the kinetic energy is calculated by summing the product of the inertia constant and the rated apparent power of all online synchronous generators. Most of the inertia constants and rated apparent powers of synchronous units are available from the Nordic grid model datasets, for the others average inertia constants have been used. For those generators modules with more than one unit, the capacity of already online units is fully utilized before considering online other units. 

The results of the simulated hydrological years have been condensed into three cases: 

\addvbuffer[4pt 6pt]{%
\setlength{\extrarowheight}{4pt}
\hskip-1.8em
\begin{tabular}{m{1pt} p{0.455\textwidth}}
     1. &  \textit{Low-inertia}: dry hydrological year, resulting in low inertia; \\
     2. &  \textit{Medium-inertia}: average hydrological year, resulting in average inertia; \\
     3. &  \textit{High-inertia}: wet hydrological year, resulting in high inertia. \\
\end{tabular}%
}
These cases have been selected because they correspond to the "best case" (high inertia), the "worst case" (low inertia) and the "most probable case" (medium inertia).

The calculated values of kinetic energy are then given as input to the frequency response model, which is explained in details in the next section. 

\section{Frequency Response Model}\label{sec:3_0}
The goal of kinetic energy estimation is to verify that the N-1 criterion is satisfied, that is to confirm that the loss of the dimensioning incident would not lead to an IFD greater than the maximum allowance. In the following, the relation between system inertia and IFD is presented, together with the statistical model used for the analyses presented in this paper.

As mentioned above, IFD depends on power deviation, system inertia and activation speed of reserves. Assuming that generators swing coherently and neglecting the frequency dependency of the load, the system dynamics can be modeled by a single machine equivalent and its behavior can be expressed using the normalized swing equation \cite{3_2}:
\begin{equation}\label{eq:3_1}
    2H \frac{d \omega_r}{dt} = P_m-P_e
\end{equation}
where $H$ is the inertia constant of the system, $\omega_r$ is the generator speed and $P_m$, $P_e$ are respectively the mechanical and electrical power of the system. The aggregate system inertia $H$ can be related to the kinetic energy $E_k$ with the following expression:
\begin{equation}\label{eq:3_3}
    H = \frac{E_k}{S_n}
\end{equation}
where $S_n$ is the system's base power. 
In such a reduced system, the rotor speed $\omega_r$ of the single machine equivalent is directly related to the system frequency $f$:
\begin{equation}\label{eq:3_4}
    \omega_r = \frac{2\pi f}{2\pi f_0}
\end{equation}											
where $f_0$ is the nominal system frequency. 
RoCoF can be obtained by plugging Eq. \eqref{eq:3_3} and \eqref{eq:3_4} into Eq. \eqref{eq:3_1}:
\begin{equation}\label{eq:3_5}
   \frac{df}{dt} = \frac{f_0}{2E_k} \Delta P
\end{equation}
where $\Delta P = (P_m-P_e)S_n$ is the mismatch between mechanical and electrical power. An expression of the IFD, $\Delta f$, can be derived by taking the Laplace transform of Eq. \eqref{eq:3_5}:
\begin{equation}\label{eq:3_6}
    \Delta f = \frac{f_0}{2s} \frac{\Delta P}{E_k}.
\end{equation}

The single machine equivalent described above can be extended including primary frequency reserves. The IFD is then expressed as in \cite{2_7}:
\begin{equation}\label{eq:3_7}
    \Delta f = \frac{\frac{f_0}{2}}{s+\frac{R\,F(s)f_0}{2E_k}}\frac{\Delta P}{E_k}
\end{equation}
with $F(s)$ the transfer function of primary reserves, describing the dynamics of governor and turbine, and $R$ the regulating strength in MW/Hz. 

Eq. \eqref{eq:3_7} shows the strong correlation between RoCoF and IFD. With the assumption that the ratio between regulating strength and kinetic energy is constant for a system with high regulating strength, the authors in \cite{2_7} approximate Eq. \eqref{eq:3_7} using a linear regression model. The regressions are expressed as:
\begin{align}
    & \Delta f_{over} \approx \alpha_{over} \frac{\Delta P}{E_k}+\beta_{over} \label{eq:3_8}\\
    & \Delta f_{under} \approx \alpha_{under} \frac{\Delta P}{E_k}+\beta_{under} \label{eq:3_9}
\end{align}
with the assumption that the transfer function of primary reserves is not the same for under and over frequency events. 

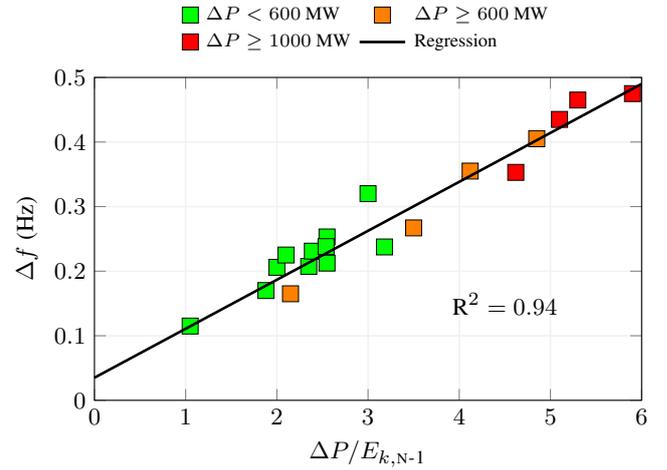
\begin{figure}[t]
    \begin{tikzpicture}
        \begin{axis}[%
                width=0.40\textwidth,
                height=0.17\textheight,
                at={(1.20in,0.806in)},
                scale only axis,
                grid=both,
                grid style={line width=0.6pt, draw=gray!10},
                xmin=0,
                xmax=6,
                xlabel style={font=\color{white!15!black}},
                xlabel={$\Delta P / E_{k,\textsc{n-1}}$},
                ymin=0,
                ymax=0.5,
                ylabel style={font=\color{white!15!black}, at={(-0.09,0.5)}},
                ylabel={$\Delta f$ (Hz)},
                ytick={0,0.1,0.2,0.3,0.4,0.5},
                axis background/.style={fill=white},
                ylabel near ticks,
                xlabel near ticks,
                legend columns=2,
                legend style={at={(0.145,1.05)}, anchor=south west, legend cell align=left, align=left, draw=none, font=\scriptsize},
                every axis legend/.append style={column sep=0.1em},
                clip mode=individual
                ]
                
                \addplot [line width=0.2pt, only marks,  mark size=3pt, mark=square*, mark options={draw=black,fill=green,solid},on layer=main] table[row sep=crcr]{%
                    1.05	0.115	\\
                    1.88	0.17	\\
                    2	    0.206	\\
                    2.1	    0.225	\\
                    2.35	0.2075	\\
                    2.39	0.231	\\
                    2.55	0.2125	\\
                    2.55	0.253	\\
                    2.54	0.238	\\
                    3	    0.32	\\
                    3.18	0.2375	\\
                };
                \addlegendentry{$\Delta P < 600$ MW}
                
                \addplot [line width=0.2pt, only marks, mark size=3pt, mark=square*, mark options={draw=black,fill=orange,solid},on layer=main] table[row sep=crcr]{%
                    2.15	0.165	\\
                    3.5	    0.267	\\
                    4.12	0.355	\\
                    4.85	0.405	\\
                };
                \addlegendentry{$\Delta P \geq 600$ MW}
                
                \addplot [line width=0.2pt, only marks, mark size=3pt, mark=square*, mark options={draw=black,fill=red,solid},on layer=main] table[row sep=crcr]{%
                    4.62	0.353	\\
                    5.1	0.435	\\
                    5.3	0.465	\\
                    5.9	0.475	\\
                };
                \addlegendentry{$\Delta P \geq 1000$ MW}
                
                \addplot [color=black, line width=1.0pt, on layer=foreground] table[row sep=crcr]{%
                    0	0.035	\\
                    6	0.49	\\
                };
                \addlegendentry{Regression}
                
                \addplot[forget plot] coordinates {(4.5, 0.15)} node[] {$\text{R}^2=0.94$};
                
        \end{axis}
    \end{tikzpicture}
    \vspace{-0.7em}
    \caption{Maximum instantaneous frequency deviation relative to the ratio between power imbalance and kinetic energy of the under-frequency regression disturbances in \cite{2_7}.}
    \label{fig:3_regression}
    \vspace{-1em}
\end{figure}

The regression analysis was carried out using respectively 19 and 26 disturbance events (occurred in the period between October 2015 and September 2016 in the Nordic system) for under and over frequency deviation, under the assumption that FCR-N were fully activated (frequency deviations start at 49.9 Hz) and provided similar response during each disturbance. \figurename~\ref{fig:3_regression} shows the 19 disturbances used to determine the under frequency regression model. The model was then validated using historical disturbances from October 2016 to September 2017, with the resulting standard deviation equal to 0.035 Hz and 0.048 Hz for under and over frequency response. This regression model is currently the most accurate model for estimating the IFD based on the kinetic energy in the NSA and it is used in Nordic TSOs' control rooms to investigate frequency instability problems \cite{2_9}.

The regression model is used in this work to determine the IFD for the 2020 and 2025 scenarios, using the kinetic energy estimations from the above mentioned market simulations. For each instance, the kinetic energy estimate is given as an input to the linear regression model (Eq. \eqref{eq:3_9}, with $\alpha_{under}=0.0757$ and $\beta_{under}=0.0369$), which returns the corresponding IFD. The IFD is then compared to the maximum allowed IFD. According to current TSOs practice, a safety margin of 0.05 Hz is kept and the maximum allowed IFD is 950 mHz. If the IFD is below this value, then the N-1 criterion is satisfied and no redispatch is necessary, so the algorithm moves to the next instance. When the IFD is greater than 950 mHz, the algorithm proceeds to compare the differnet mitigation strategies considered. 
\section{Remedial Actions: Technical Considerations}\label{sec:4}
Remedial actions are defined as the set of measures applied by TSOs to maintain operational security and relieve congestions. According to the Network Code on System Operation \cite{2_19}, remedial actions can be divided into:

\addvbuffer[4pt 6pt]{%
\setlength{\extrarowheight}{2pt}
\hskip-1.5em
\begin{tabularx}{0.488\textwidth}{l *1{>{\arraybackslash}X}}
\vspace{-0.1em}
     -- & \textit{Preventive actions}: measures applied in operational planning or scheduling stage to prevent dangerous situations and maintain system security in the coming operational situation. \\
     -- & \textit{Corrective actions}: measures implemented immediately or relatively soon after an occurrence of a contingency.
\end{tabularx}%
}

In this paper we consider two remedial actions taken to meet the maximum allowed IFD during low inertia periods, fulfilling the \mbox{N-1} security criterion. This section starts with the explanation of the current paradigm, the \textit{preventive} reduction of O3 followed by upward regulation of reserves. We then investigate an alternative \textit{corrective} action where HVDC contributes to frequency stability. For the sake of completeness, we conclude this section discussing in brief other remedial actions.

\vspace{-0.3em}
\subsection{Current Paradigm - DI Reduction}\label{ssec:4_1}
The current practice in case of low inertia periods is the \textit{preventive} reduction of the dimensioning incident, i.e. the largest production unit, O3. For example, if the maximum disconnected power the system can handle is 1'300 MW, the power output of O3 is reduced by 150 MW. 

According to TSOs' current practice, the dimensioning incident is usually reduced by blocks of 50 MW. Moreover, a certain security margin is kept and the redispatch is performed few hours before and after the low-inertia event. Due to technical limitations, nuclear units might need few hours to ramp down (normally 8 hours) and, depending on the length of the reduction period, they might take from 6 to 72 hours to get back to their nominal power output (for example, if the limitation is performed for up to 80\% of the operational period, the output cannot be increased for the remaining time) \cite{2_9}. For these reasons, if the frequency limits are exceeded twice (or more) within 36 hours, the dimensioning incident is reduced for the whole period (and the maximum reduction is applied).

\figurename~\ref{fig:3_power} shows the power limitation on O3 (orange plot) depending on the level of inertia. These values are obtain through dynamic simulations using a single machine equivalent model of the Nordic power system. The parameters of the model are tuned based on the frequency response of the system to some real disturbances happened in the Nordic system. The model has been validated and is currently used by Nordic TSOs in their control rooms. For this analysis, the reference incident is applied with varying system inertia and the IFD is calculated. The model considers that FCR-N are already fully activated and the frequency deviation is calculated starting from 49.9 Hz. With an IFD greater than 0.9 Hz, the reference incident is decreased until the frequency limits are not exceeded anymore.

\vspace{-0.4em}
\subsection{HVDC Emergency Power Control}\label{ssec:4_3}
As electricity markets are becoming increasingly integrated towards a common European market, TSOs have started exchanging balancing energy products through common European platforms. With this purpose, new cooperation projects have been kicked off. For example, the International Grid Control Cooperation (IGCC) project aims at minimizing the activation of aFRR in neighboring control areas in case of imbalances in opposite directions \cite{4_2}.

In a similar fashion, TSOs could use HVDC links for sharing FCR-D, decreasing the risk of frequency violations during low inertia periods. Indeed, HVDC converters, equipped with fast frequency controllers, can adjust the power flow in response to frequency deviations. In the Nordic countries, this control mode is referred to as Emergency Power Control (EPC). This measure falls in the category of \textit{corrective} actions: in low inertia periods, although the dimensioning incident would lead to an IFD greater than 1 Hz, the output of O3 is not reduced in advance. In case the dimensioning incident occurs, EPC is immediately activated and the power necessary to keep the frequency within the limits is injected through HVDC. 

\definecolor{Blue1}{rgb}{0.14510,0.23137,0.27843}%
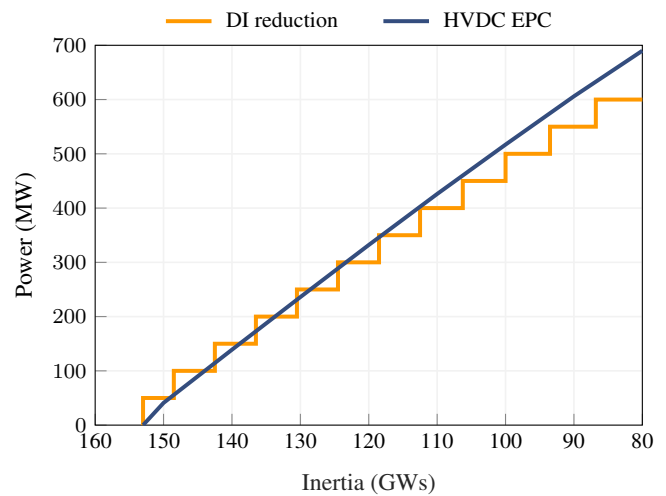
\begin{figure}
    \centering
    \begin{tikzpicture}
        \vspace{-0.5em}
        \begin{axis}[%
            width=0.4\textwidth,
            height=0.2\textheight,
            at={(0.758in,0.481in)},
            scale only axis,
            grid=both,
            grid style={line width=0.6pt, draw=gray!10},
            xmin=-160,
            xmax=-80,
            xtick pos=left,
            ytick pos=left,
            xlabel style={font=\color{white!15!black}},
            xlabel={Inertia (GWs)},
            ymin=0,
            ymax=700,
            ytick = {0,100,200,300,400,500,600,700},
            xtick = {-160,-150,-140,-130,-120,-110,-100,-90,-80},
            xticklabels = { {160}, {150}, {140}, {130}, {120}, {110}, {100}, {90}, {80}},
            every tick label/.append style={font=\footnotesize},
            ylabel style={font=\color{white!15!black}},
            ylabel={ Power (MW)},
            ylabel near ticks,
            axis background/.style={fill=white},
            legend columns=2,
            legend style={at={(0.11,1.01)}, anchor=south west, legend cell align=left, align=left, draw=none, font=\footnotesize},
            /tikz/every even column/.append style={column sep=0.5cm},
            every axis legend/.append style={column sep=0.3em},
            clip mode=individual
            ]
            \addplot [color=orange1, line width=1.5pt]%, mark size=2pt, mark=square*, mark options={draw=orange1,fill=white,solid,line width=1pt}]
            table[row sep=crcr]{%
                % -80	    600\\
                % -90	    527\\
                % -100	450\\
                % -110	370\\
                % -120	289\\
                % -130	205\\
                % -140	121\\
                % -150	36\\
                % -153	0\\
                -153	0	\\
                -153	50	\\
                -148.5	50	\\
                -148.5	100	\\
                -142.5	100	\\
                -142.5	150	\\
                -136.5	150	\\
                -136.5	200	\\
                -130.5	200	\\
                -130.5	250	\\
                -124.5	250	\\
                -124.5	300	\\
                -118.5	300	\\
                -118.5	350	\\
                -112.5	350	\\
                -112.5	400	\\
                -106.25	400	\\
                -106.25	450	\\
                -100	450	\\
                -100	500	\\
                -93.5	500	\\
                -93.5	550	\\
                -86.8	550	\\
                -86.8	600	\\
                -80	600	\\
            };
            \addlegendentry{DI reduction}
            
            \addplot [color=blue1, line width=1.5pt]%, mark size=2pt, mark=square*, mark options={draw=blue1,fill=white,solid,line width=1pt}]
            table[row sep=crcr]{%
                -80	    690\\
                -90	    606\\
                -100	517\\
                -110	426\\
                -120	332\\
                -130	236\\
                -140	139\\
                -150	41\\
                -153	0\\
            };
            \addlegendentry{HVDC EPC}

        \end{axis}
    \end{tikzpicture}%
    \vspace{-0.5em}
    \caption{Required DI reduction and injected power by HVDC EPC, respectively, to maintain the N-1 security criterion in the Nordic system with low inertia.}
    \label{fig:3_power}
    \vspace*{-1.3em}
\end{figure}

Different control strategies can be used to define the response of HVDC converters. The currently implemented strategy is based on step-wise triggers: depending on the size of the power deviation and the corresponding frequency variation, a constant amount of power is injected to improve the frequency response of the system. %\figurename~\ref{fig:4_nadir} shows how the frequency nadir (the lowest value that the frequency reaches after a contingency) is increased using EPC, and what is the equivalent amount of power injected depending on the level of inertia of the system \cite{2_9, 2_13}.
Authors in \cite{2_13} presented a new approach based on droop control, where the power injection varies taking into consideration the actual frequency response of the system (instead of injecting a fixed amount of power) to achieve a faster response compared to the current strategy. However, detailed technical analyses of such control strategies are outside the scope of this paper and for the comparison of the two remedial actions the current implemented strategy (step-wise trigger) is considered.

Similar to the analysis done for DI reduction, the necessary power injected through HVDC links depending on the inertia level of the system is plotted in \figurename~\ref{fig:3_power} (blue plot). The analysis is carried out with the single machine equivalent model described above. Similar to the work in \cite{2_13}, the frequency response of the system is analyzed with varying system inertia: the reference incident is applied and the frequency nadir is calculated. The contribution of HVDC links to frequency stability is assessed considering that all HVDC links have a single-step injection, the same triggering level and the same time threshold. As pointed out by the authors in \cite{2_13}, the triggering level has a great impact on the system response: the greater the frequency dead band, the more power must be injected. Our analysis are carried out using the frequency dead band values provided in \cite{2_9}. It is no surprise that more power shall be injected through HVDC compared to the reduction of the incident, as the frequency dead band makes the contribution of HVDC links less effective. However, reservation of HVDC capacity and procurement of reserves for HVDC EPC are only needed for those hours when the frequency can fall below 49.05 Hz whereas the reduction of the dimensioning incident would be prolonged for more hours due to technical limitations.

The cost saving analysis focuses on four interconnectors - Kontek (KO), Baltic Cable (BC), NorNed (NN) and SwePol (SP) - and the injected power is equally shared by the four links. Since TSOs must guarantee the security of the system, available transmission capacities given to the market already consider what remedial actions are taken in case of dimensioning incident. Moreover, this remedial action is used only to contain the frequency within the limit while the frequency restoration is assumed to use local reserves. It follows that the power injections last for a very short period, in the range of minutes, during which the thermal limits of AC cables can be exceeded to some extent. Thus, a possible congestion of the AC system does not prevent the utilization of any remedial action, in particular HVDC EPC.

The ongoing standardization of frequency stability services and unification of ancillary service markets leave the door open to new agreements between balancing entities for improving the security of the system while reducing the costs for energy consumers. In this frame, the utilization of HVDC links for sharing FCR-D or FFR could represent the first joint effort to secure different synchronous areas and prepare the system to large frequency deviations during low inertia periods. Indeed, this is the main difference with the IGCC project, as the latter aims at reducing the costs related to unnecessary activation of frequency reserves, while the purpose of HVDC EPC is to provide additional support to local FCR-D when necessary.

\vspace{-0.4em}
\subsection{Other Remedial Actions}\label{ssec:4_4}
When considering other mitigation strategies, these can be distinguished according to the time frame of their applicability. A series of short-term strategies could be the procurement of additional FCR-D, the utilization of certain thermal and hydro power units as synchronous compensators, the installation of new synchronous condensers and the decrease of non-synchronous generation, imports from HVDC connections and loads \cite{2_9}. 

Among the medium-term strategies, new market products for frequency support, e.g. Fast Frequency Reserves (FFR), are probably the most interesting solution \cite{3_2a}. Another option is to adjust the existing controllers of non-synchronous-connected production units for providing synthetic inertia \cite{3_2b}. 

Finally, the adjustment of protection schemes reacting to high RoCoF and low frequencies \cite{2_10j}, the implementation of inertia constraints in market or dispatch algorithms and the introduction of new inertia products \cite{3_2c, 3_2d} can be seen as long-term strategies.

All these remedial actions have their benefits and drawbacks. Their costs vary greatly depending on the system under consideration, thus the comparison of these actions is case sensitive. In the next section, these costs will be briefly discussed.

\section{Remedial Actions: Market Considerations}\label{sec:5}
In this section, the market considerations and price scenarios for the cost saving analyses are presented.

\definecolor{BlueREG2}{rgb}{0.1804,0.4588,0.7137}%
\definecolor{BlueREG}{rgb}{0.4000,0.6667,0.8431}%
\definecolor{OrangeREG}{rgb}{0.9294,0.4902,0.1922}%

\pgfplotsset{
boxplot/every median/.style={
/tikz/color=red,
},
}
\begin{figure}[!b]
\vspace{1.2em}
    \begin{tikzpicture}
        \draw[color=OrangeREG,line width = 0.7pt,dashed] (3.78,1.78) -- (4.32,0);
        \draw[color=OrangeREG,line width = 0.7pt,dashed] (3.78,1.92) -- (4.32,4.3);
        
        \begin{axis}[%
            width=0.21\textwidth,
            height=0.17\textheight,
            at={(0,0)},%{(1.20in,0.806in)},
            scale only axis,
            axis y line=none,
            axis x line=none,
            xmin=0,
            xmax=7,
            xtick={},
            ymin=0,
            ymax=80,
            ytick={},
            axis background/.style={fill=white}
            ]
            
            \addplot[xbar interval, fill=OrangeREG, fill opacity= 0.2, draw=none, line width = 0.3pt, bar width = 10pt] table[row sep=crcr] {%
                x	y	\\
                7	33.24	\\
                7   35.92 \\
                };
        \end{axis}      
            
        \begin{axis}[%
            width=0.21\textwidth,
            height=0.17\textheight,
            at={(0,0)},%{(1.20in,0.806in)},
            scale only axis,
            boxplot/draw direction=y,
            xmin=0,
            xmax=7,
            xtick={1,2,3,4,5,6},
            xticklabels={{2015},{2016},{2017},{2018},{2019},{5-year}},
            xticklabel style={rotate=90},
            ymin=0,
            ymax=80,
            ylabel style={font=\color{white!15!black}},
            ylabel={Price (\euro/MWh)},
            %axis background/.style={fill=none},
            ylabel near ticks,
            xlabel near ticks
            ]
 
            \addplot+ [boxplot prepared={
                lower whisker=1.21, lower quartile=11.97,
                median=17.38, upper quartile=22.65,
                upper whisker=38.67},
                draw=BlueREG,fill=white,line width = 0.5pt] coordinates {};
            
            \addplot+ [boxplot prepared={
                lower whisker=16.58, lower quartile=25.76,
                median=29.10, upper quartile=31.95,
                upper whisker=41.28},
                draw=BlueREG,fill=white,line width = 0.5pt] coordinates {};
            
            \addplot+ [boxplot prepared={
                lower whisker=16.59, lower quartile=27.53,
                median=31.29, upper quartile=35.72,
                upper whisker=48.13},
                draw=BlueREG,fill=white,line width = 0.5pt] coordinates {};
            
            \addplot+ [boxplot prepared={
                lower whisker=36.88, lower quartile=49.46,
                median=54.09, upper quartile=58.42,
                upper whisker=71.74},
                draw=BlueREG,fill=white,line width = 0.5pt] coordinates {};
                
            \addplot+ [boxplot prepared={
                lower whisker=20.38, lower quartile=33.10,
                median=37.34, upper quartile=41.45,
                upper whisker=54.09},
                draw=BlueREG,fill=white,line width = 0.5pt] coordinates {};
            
            \addplot+ [boxplot prepared={
                lower whisker=1.21, lower quartile=25.56,
                median=32.28, upper quartile=43.13,
                upper whisker=71.74},
                draw=BlueREG,fill=white,line width = 1pt, solid] coordinates {};
                
        \end{axis}
        
        \begin{axis}[%
            width=0.15\textwidth,
            height=0.17\textheight,
            at={(160,0)},
            axis line style={OrangeREG, line width = 0.7pt},
            ytick style={OrangeREG},
            xtick style={OrangeREG},
            axis on top,
            scale only axis,
            yticklabel pos=right,
            ymin=33.55,
            ymax=36,
            xmin=0,
            xmax=10,
            xlabel={Probability (\%)},
            axis background/.style={fill=OrangeREG, fill opacity = 0.1},
            xlabel near ticks
            ]
            
            \addplot[xbar interval, fill=BlueREG, fill opacity=0.6, draw=BlueREG] table[row sep=crcr] {%
                x	y	\\
                0.001	33.24	\\
                0.002	33.307	\\
                0.005	33.374	\\
                0.012	33.441	\\
                0.023	33.508	\\
                0.044	33.575	\\
                0.095	33.642	\\
                0.154	33.709	\\
                0.306	33.776	\\
                0.575	33.843	\\
                0.844	33.91	\\
                1.382	33.977	\\
                2.052	34.044	\\
                2.803	34.111	\\
                3.815	34.178	\\
                5.098	34.245	\\
                6.26	34.312	\\
                7.165	34.379	\\
                8.042	34.446	\\
                8.507	34.513	\\
                8.569	34.58	\\
                8.338	34.647	\\
                7.89	34.714	\\
                6.856	34.781	\\
                5.752	34.848	\\
                4.586	34.915	\\
                3.431	34.982	\\
                2.597	35.049	\\
                1.817	35.116	\\
                1.164	35.183	\\
                0.754	35.25	\\
                0.462	35.317	\\
                0.27	35.384	\\
                0.158	35.451	\\
                0.094	35.518	\\
                0.043	35.585	\\
                0.018	35.652	\\
                0.011	35.719	\\
                0.004	35.786	\\
                0.001	35.853	\\
                0.001	35.92	\\
            };
                
            \addplot [color=red,dashed, line width =0.3pt] table[row sep=crcr]{%
                0	34.6	\\
                10   34.6	\\
            };
            
            \addplot [color=BlueREG2,dashed, line width = 0.3pt] table[row sep=crcr]{%
                0	34.1	\\
                10   34.1	\\
            };
            
            \addplot [color=BlueREG2,dashed,line width = 0.3pt] table[row sep=crcr]{%
                0	35.1	\\
                10   35.1	\\
            };
            
        \end{axis}
    \end{tikzpicture}
    \vspace{-0.3em}
    \caption{Left: Nordic regulating power prices during the summer period (calculated for every hour as the average price in the 12 bidding zones). Right: distribution of the average price obtained after data resampling. Blue dashed lines are respectively the 5th and 95th percentiles, while the red one is the median.}
    \label{fig:5_regprices}
    \vspace{-1em}
\end{figure}
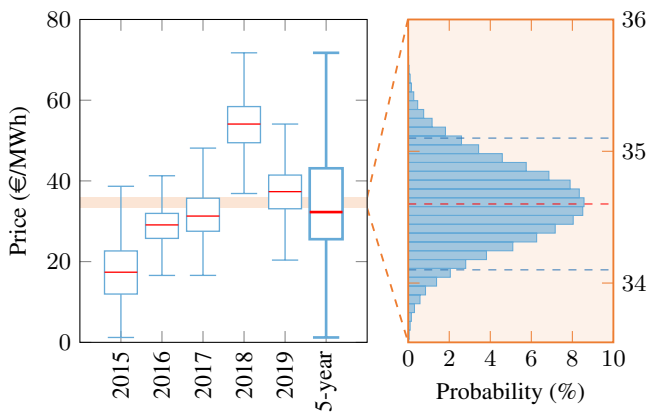

\subsection{Redispatching Costs and Regulating Prices}\label{ssec:5_1}

The downregulation of O3 is considered as a redispatching action. Normally, redispatch happens after the day-ahead and intra-day markets have been cleared: this is done to avoid the distortion of the market outcome. Generators and consumers have to submit their final dispatch 45 minutes before real time operation; in this time frame, TSOs check if the actual dispatch violates grid constraints. If this happens, they downregulate and upregulate some units. In the Nordic countries, this is done following a market-based approach: generators and consumers submit their bids for up/down regulation, the real-time market is cleared and the prices for up/down regulation are defined \cite{4_5}. Although each Nordic TSO is responsible for the real-time power balance in their control zone, there is a common regulating power market operated by Nord Pool. This is done to reduce balancing costs and to increase the competition between balancing responsible entities. 

In case of low inertia periods, the security of the system is considered in danger and a different regulation applies: Svk can communicate the limitation to OKG at any market stage. When this measure is used, the producer should receive market compensation for the costs associated with the power limitation. First, by decreasing its power output, the producer incurs opportunity costs that are equal to what they would have received for producing an amount of power equal to the power reduction. Second, by moving away from the nominal power output, extra costs are incurred due to lower efficiency (as a rule of thumb, for nuclear power plants, one can say that half of the fuel which is not used during the power reduction is lost and cannot be used later on) \cite{2_9}. Third, the decrease of power production of nuclear power plants results in a temperature transient, inducing a cumulative aging of the unit and increasing the risk of failure \cite{2_9}. All these things are taken into consideration by Nordic TSOs and OKG receives a financial compensation, as stated on a bilateral agreement between Svk and OKG.

In this paper, we consider that low-inertia events are forecast after the day-ahead market is cleared, and thus the reduction of O3 is performed similarly to normal redispatching:

\addvbuffer[4pt 6pt]{%
\setlength{\extrarowheight}{2pt}
\hskip-1.5em
\begin{tabularx}{0.488\textwidth}{l *1{>{\arraybackslash}X}}
\vspace{-0.1em}
     -- & OKG is compensated for the opportunity cost of not producing 100 MW (the compensation was equal to 49 SEK/MWh - approx. 4.64 \euro/MWh); \\
     -- & OKG is compensated for reduced efficiency and other costs associated with the power limitation (fixed amount equal to 50'000 SEK - approx. 4'740 \euro); \\
     -- & The substitute power is procured form other generators in the regulating market (Nord Pool regulating price \euro/MWh \cite{4_1}).
\end{tabularx}%
}

\begin{table}[!b]
%\vspace{-1em}
    \caption{Average prices for 2018 and for the period 2015-2019.}
    %\vspace{-0.3em}
    \label{tab1:prices}
    \small
    \centering
        \begin{tabularx}{0.47\textwidth}{p{2.2cm} *1{>{\centering\arraybackslash}X} cccc}% *3{>{\centering\arraybackslash}X}}p{1.5cm} *1{>{\arraybackslash}X}
        \hline
                    &     & \textbf{2018}   & \multicolumn{3}{c}{\textbf{5-year average}}     \Tstrut\\
                    &     &       & {5th pct} & {Median} & {95th pct}          \Bstrut\\ 
        \hline
        \multirow{2}{*}{\shortstack[l]{Regulating power$\,\,$ \\ (\euro/MWh)}} &  &  \multirow{2}{*}{54.06} & \multirow{2}{*}{34.10} & \multirow{2}{*}{34.60} & \multirow{2}{*}{35.10} \Tstrut\Bstrut\\
        & & & & & \\
        \hline
        \multirow{3}{*}{\shortstack[l]{FCR procurement$\,\,$ \\ (\euro/MW)}} 	&	DE	&	11.18	&	9.61	&	10.78	&	12.13	\Tstrut\\
        	&	NL	&	19.53	&	12.15	&	13.63	&	\text{15.19 }	\\
        	&	PL	&	5.34	&	5.32	&	5.34	&	5.36	\Bstrut\\
        \hline
        \multirow{4}{*}{\shortstack[l]{Congestion rent$\,\,$ \\ (\euro/MWh)}}	&	KO	&	1.27	&	5.27	&	5.59	&	5.93	\Tstrut\\
        	&	BC	&	1.78	&	6.60	&	6.94	&	\text{7.30 }	\\
        	&	NN	&	5.01	&	11.57	&	11.97	&	\text{12.39 }	\\
        	&	SP	&	2.00	&	10.75	&	11.31	&	11.91	\Bstrut\\
        \hline
    \end{tabularx}
\end{table}

Although there might be some small changes in the bilateral agreement between OKG and Svk (a new version is expected for 2020), we calculate the financial compensation in the 2020 and 2025 scenarios using the values from 2018. 

Concerning the regulating power prices, all three low-inertia events in 2018 happened during summer (June to August). Therefore, in all our scenarios we assume that the redispatching will most probably occur in the summer period. The distribution of regulating prices for the last 5 years is shown in \figurename~\ref{fig:5_regprices} (left): prices are calculated as the average price in the 12 Nordic bidding zones during a specific hour \cite{4_1}. Since market forecasts for the year 2020 or 2025 are out of the scope of this work, reasonable prices are calculated through data re-sampling. \figurename~\ref{fig:5_regprices} (right) shows the distribution of the average price obtained by randomly sampling 2'232 prices out of the 11'160 considered (this action is repeated 10'000 times). From the distribution, three price scenarios (high-medium-low) are obtained considering the 5th, 50th (median) and the 95th percentile. These prices, as well as the average price for summer 2018, are reported in \tablename~\ref{tab1:prices}.

It is important to notice that, if the downregulation is performed before the day-ahead market closure, the out-of-pocket costs incurred by TSOs would be lower since the substitute power is supplied by other generators within the market operation. However, this happens at the expense of society, as more expensive generators will produce this power. If there is an increment in day-ahead prices, all the consumers are subject to socio-economic loss.

\vspace{-0.3em}
\subsection{HVDC Reservation and FCR Procurement Costs}\label{ssec:5_2}
The utilization of HVDC lines for frequency support relies on the fact that there is enough transmission capacity available on the HVDC interconnectors, and that there is a certain availability of frequency reserves on the neighboring systems. As for now, there seems to be no regulation about how these can be procured on a market basis. In order to assess what could be the cost of this remedial action, we envision a possible future situation where there is a European market for reserves, and Nordic TSOs are requested to procure the necessary primary reserves through this platform. In fact, this seems to be the direction that European countries are taking, as described in \cite{4_6} for automatic activated FRR. Moreover, the reservation of HVDC capacity is assumed to come with a cost. This is considered also in \cite{2_9}, where they assume there might be a reservation cost for HVDC in the future.

However, it cannot be excluded that Nordic TSOs might sign an agreement - similar to IGCC - with neighboring TSOs for the exchange of reserves in situations where operational security is in danger (or at any time in order to increase the level of security); or that HVDC capacity will not be reserved at a cost, as there are 100 MW of available capacity on the considered lines for 70\% of the time on a yearly average \cite{4_1} or, alternatively, HVDC lines could be overloaded for a short amount of time (in the range of minutes) \cite{4_1a, 4_1b}. Despite these considerations, the analyses presented in this paper consider that these services will be provided at a cost. 

\definecolor{GreenFCR1}{rgb}{0.19608,0.29020,0.07451}%
\definecolor{GreenFCR2}{rgb}{0.46667,0.67451,0.18824}%
\definecolor{GreenFCR3}{rgb}{0.76863,0.87843,0.13333}%

\begin{figure}[t]
    \begin{tikzpicture}
        \begin{axis}[%
            width=0.15\textwidth,
            height=0.17\textheight,
            at={(0,0)},
            scale only axis,
            axis on top,
            ymin=0,
            ymax=15,
            xmin=5.28,
            xmax=5.42,
            ylabel={Probability (\%)},
            axis background/.style={white},
            xlabel near ticks,
            ylabel near ticks,
            legend columns=1,
            legend style={at={(0.2,1.025)}, anchor=south west, legend cell align=left, align=left, draw=none, font=\footnotesize},
            every axis legend/.append style={column sep=0.1em}
            ]
            
            \addplot[ybar interval, fill=GreenFCR2, fill opacity=0.6, draw=GreenFCR2, area legend] table[row sep=crcr] {%
                x	y	\\
                5.286	0.001666528	\\
                5.2894	0	\\
                5.2928	0.011665695	\\
                5.2962	0.01499875	\\
                5.2996	0.069994167	\\
                5.303	0.136655279	\\
                5.3064	0.339971669	\\
                5.3098	0.816598617	\\
                5.3132	1.48487626	\\
                5.3166	2.616448629	\\
                5.32	3.863011416	\\
                5.3234	5.957836847	\\
                5.3268	7.871010749	\\
                5.3302	9.845846179	\\
                5.3336	11.26572786	\\
                5.337	11.67902675	\\
                5.3404	11.19406716	\\
                5.3438	9.874177152	\\
                5.3472	7.949337555	\\
                5.3506	5.799516707	\\
                5.354	3.863011416	\\
                5.3574	2.446462795	\\
                5.3608	1.40988251	\\
                5.3642	0.783268061	\\
                5.3676	0.383301392	\\
                5.371	0.176651946	\\
                5.3744	0.088325973	\\
                5.3778	0.024997917	\\
                5.3812	0.01499875	\\
                5.3846	0.008332639	\\
                5.388	0.008332639	\\
            };
                
            \addplot [color=red,dashed, line width =0.6pt, forget plot] table[row sep=crcr]{%
                5.34	0	\\
                5.34   15	\\
            };
            \legend{Poland}
        \end{axis}
        
        \begin{axis}[%
            width=0.23\textwidth,
            height=0.17\textheight,
            at={(73,0)},
            scale only axis,
            axis on top,
            ymin=0,
            ymax=15,
            xmin=7.6,
            xmax=18.2,
            yticklabels={},
            xtick={{8},{10},{12},{14},{16},{18}},
            xlabel={Price (\euro/MW)},
            xlabel style={font=\color{white!15!black},xshift=-4.8em},
            axis background/.style={white},
            ylabel near ticks,
            legend columns=2,
            legend style={at={(-0.15,1.025)}, anchor=south west, legend cell align=left, align=left, draw=none, font=\footnotesize},
            every axis legend/.append style={column sep=0.1em}
            ]
            
            \addplot[ybar interval, fill=GreenFCR1, fill opacity=0.6, draw=GreenFCR1, area legend] table[row sep=crcr] {%
                x	y	\\
                8.1	0.002443942	\\
                8.26	0.004887884	\\
                8.42	0.017107595	\\
                8.58	0.063542494	\\
                8.74	0.141748641	\\
                8.9	0.347039775	\\
                9.06	0.624427201	\\
                9.22	1.037453412	\\
                9.38	1.638663164	\\
                9.54	2.670006721	\\
                9.7	3.658581292	\\
                9.86	4.836561373	\\
                10.02	5.871570844	\\
                10.18	6.918800024	\\
                10.34	7.550559052	\\
                10.5	8.247082544	\\
                10.66	8.325288691	\\
                10.82	8.209201442	\\
                10.98	7.562778762	\\
                11.14	6.58397996	\\
                11.3	5.946111077	\\
                11.46	4.941650883	\\
                11.62	4.036170343	\\
                11.78	2.959613857	\\
                11.94	2.346184395	\\
                12.1	1.759638297	\\
                12.26	1.248854402	\\
                12.42	0.837050162	\\
                12.58	0.625649172	\\
                12.74	0.380032993	\\
                12.9	0.230952526	\\
                13.06	0.15274638	\\
                13.22	0.103867538	\\
                13.38	0.052544755	\\
                13.54	0.030549276	\\
                13.7	0.019551537	\\
                13.86	0.008553797	\\
                14.02	0.004887884	\\
                14.18	0.001221971	\\
                14.34	0.001221971	\\
                14.5	0.001221971	\\
            };
            
            \addplot [color=red,dashed, line width =0.6pt, forget plot] table[row sep=crcr]{%
                10.78	0	\\
                10.78   15	\\
            };

            \addplot[ybar interval, fill=GreenFCR3, fill opacity=0.6, draw=GreenFCR3, area legend] table[row sep=crcr] {%
                x	y	\\
                10.1	0.00499985	\\
                10.272	0.00599982	\\
                10.444	0.00699979	\\
                10.616	0.01299961	\\
                10.788	0.04899853	\\
                10.96	0.10499685	\\
                11.132	0.17699469	\\
                11.304	0.34698959	\\
                11.476	0.50798476	\\
                11.648	0.869973901	\\
                11.82	1.268961931	\\
                11.992	1.733947982	\\
                12.164	2.514924552	\\
                12.336	3.223903283	\\
                12.508	3.972880814	\\
                12.68	4.863854084	\\
                12.852	5.750827475	\\
                13.024	6.521804346	\\
                13.196	6.964791056	\\
                13.368	7.343779687	\\
                13.54	7.412777617	\\
                13.712	7.103786886	\\
                13.884	6.897793066	\\
                14.056	6.279811606	\\
                14.228	5.559833205	\\
                14.4	4.711858644	\\
                14.572	3.994880154	\\
                14.744	3.141905743	\\
                14.916	2.494925152	\\
                15.088	1.835944922	\\
                15.26	1.395958121	\\
                15.432	1.001969941	\\
                15.604	0.673979781	\\
                15.776	0.46098617	\\
                15.948	0.29599112	\\
                16.12	0.19099427	\\
                16.292	0.12899613	\\
                16.464	0.06899793	\\
                16.636	0.0399988	\\
                16.808	0.02599922	\\
                16.98	0.01399958	\\
                17.152	0.00599982	\\
                17.324	0.00799976	\\
                17.496	0.00199994	\\
                17.668	0.00299991	\\
                17.84	0.00299991	\\
            };
                
            \addplot [color=red,dashed, line width =0.6pt, forget plot] table[row sep=crcr]{%
                13.73	0	\\
                13.73   15	\\
            };
            \legend{Germany, Netherlands}
        \end{axis}

    \end{tikzpicture}
    \vspace{-0.3em}
    \caption{Distribution of the average price for primary reserves in Germany, the Netherlands and Poland obtained after data re-sampling.}
    \label{fig:6_fcrprices}
    \vspace{-1em}
\end{figure}
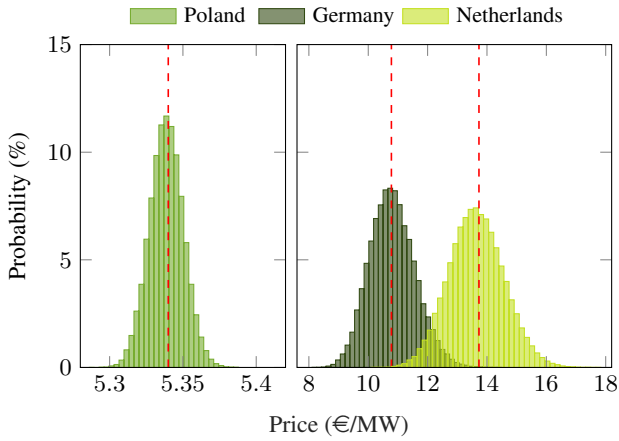

Regarding FCR procurement prices in Germany, the Netherlands and Poland, this data can be found in the ENTSO-E Transparency Platform \cite{4_7}. Similar to what was done for regulating power prices, prices corresponding to one summer period were sampled from the data set of the summer prices of the last 5 years, and three price scenarios (high-medium-low) were determined using the 5th, 50th (median) and 95th percentile, shown in \tablename~\ref{tab1:prices} together with the average summer prices of 2018. The distribution of the average prices for these three countries is depicted in \figurename~\ref{fig:6_fcrprices}.

The only source of prices for HVDC capacity is Energinet's data platform: the Energi Data Service \cite{4_4}. In their web page, the yearly and monthly auction prices for Kontek, the link between Germany and Denmark (DK2), can be found. However, there is no data for Baltic Cable, NorNed and SwePol. Since TSOs do not incur direct costs when reserving transmission capacity, this should be provided at the opportunity cost of not using it for energy trades in the day-ahead market. For this reason, in this paper we consider that the price for reserving HVDC capacity in a specific hour is equal to the congestion rent (in that hour). Day-ahead spot prices can be found in the ENTSO-E Transparency Platform. Similar to the other prices, congestion rents corresponding to one summer period were sampled from the data set of the last 5 summers. Again, three prices (high-medium-low) were determined using the 5th, 50th (median) and 95th percentile, shown in \tablename~\ref{tab1:prices} together with the average congestion rent of summer 2018. The distribution of the average prices for these four interconnectors is depicted in \figurename~\ref{fig:7_hvdcprices}. The resulting prices for Kontek were compared to the actual auction prices, with the calculated prices slightly higher than the actual ones.

\vspace{-0.4em}
\subsection{Costs of other Remedial Actions}\label{ssec:5_3}
In the following, the potential costs of some of the other mitigation strategies introduced in Section \ref{ssec:4_4} are shortly discussed.

Similar to DI reduction, when the output of non-synchronous-connected units is decreased to increase synchronous generation, power producers must receive compensation for the opportunity cost of not producing that energy. The socioeconomic cost of this remedial action is thus equal to the fuel cost of the replacing synchronous units. 

In a similar way, if HVDC imports are decreased, the cost is equal to the loss of congestion rent. Moreover, if the HVDC capacity given to the market is reduced in advance, this might have an impact on power prices and those market participants which are subject to higher prices incur extra costs. 

The disconnection of industrial or aggregated loads, instead, requires additional control equipment for remote activation and control. Moreover, new agreements with consumers must be signed, resulting in new flexibility contracts with financial compensation for the energy not supplied. Finally, for most of industrial loads, the disruption of electricity might results in additional activation costs once industrial processes are started up again.

\color{black}
% \definecolor{PinkHVDC}{rgb}{0.87843,0.62353,0.78039}%
% \definecolor{BlueHVDC1}{rgb}{0.08235,0.26275,0.38039}%
% \definecolor{BlueHVDC2}{rgb}{0.54510,0.72941,0.85098}%
% \definecolor{PurpleHVDC}{rgb}{0.49400,0.18400,0.55600}%
\definecolor{Brown1HVDC}{rgb}{0.7255,0.6824,0.5059}%
\definecolor{Brown2HVDC}{rgb}{0.5216,0.3529,0.3176}%
\definecolor{YellowHVDC}{rgb}{1,0.7529,0}%
\definecolor{OrangeHVDC}{rgb}{0.9294,0.4902,0.1922}%

\begin{figure}[t]
    \begin{tikzpicture}
        \begin{axis}[%
            width=0.39\textwidth,
            height=0.17\textheight,
            at={(0,0)},
            scale only axis,
            axis on top,
            ymin=0,
            ymax=15,
            xmin=4,
            xmax=14,
            ylabel={Probability (\%)},
            xlabel={Congestion rent (\euro/MWh)},
            axis background/.style={white},
            xlabel near ticks,
            ylabel near ticks,
            legend columns=4,
            legend style={at={(-0.03,1.025)}, anchor=south west, legend cell align=left, align=left, draw=none, font=\footnotesize},
            every axis legend/.append style={column sep=0.1em}
            ]
            
            \addplot[ybar interval, fill=YellowHVDC, fill opacity=0.6, draw=YellowHVDC, area legend] table[row sep=crcr] {%
                x	y	
                4.68	0.00099999	\\
                4.726	0	\\
                4.772	0.00299997	\\
                4.818	0.00399996	\\
                4.864	0.01599984	\\
                4.91	0.0399996	\\
                4.956	0.07499925	\\
                5.002	0.13699863	\\
                5.048	0.29799702	\\
                5.094	0.53899461	\\
                5.14	0.92899071	\\
                5.186	1.50798492	\\
                5.232	2.15597844	\\
                5.278	3.31396686	\\
                5.324	4.50295497	\\
                5.37	5.733942661	\\
                5.416	6.900930991	\\
                5.462	8.011919881	\\
                5.508	8.765912341	\\
                5.554	9.068909311	\\
                5.6	8.899911001	\\
                5.646	8.328916711	\\
                5.692	7.355926441	\\
                5.738	6.313936861	\\
                5.784	5.072949271	\\
                5.83	3.95696043	\\
                5.876	2.83897161	\\
                5.922	1.96198038	\\
                5.968	1.32598674	\\
                6.014	0.81399186	\\
                6.06	0.48399516	\\
                6.106	0.30199698	\\
                6.152	0.16299837	\\
                6.198	0.09399906	\\
                6.244	0.04499955	\\
                6.29	0.01699983	\\
                6.336	0.01499985	\\
                6.382	0.00299997	\\
                6.428	0.00099999	\\
                6.474	0.00099999	\\
                6.52	0.00099999	\\
            };

            \addplot[ybar interval, fill=OrangeHVDC, fill opacity=0.6, draw=OrangeHVDC, area legend] table[row sep=crcr] {%    		
                x	y	
                6.08	0.00099999	\\
                6.127	0.00699993	\\
                6.174	0.01599984	\\
                6.221	0.03799962	\\
                6.268	0.07199928	\\
                6.315	0.13799862	\\
                6.362	0.27499725	\\
                6.409	0.49299507	\\
                6.456	0.81799182	\\
                6.503	1.27498725	\\
                6.55	1.92898071	\\
                6.597	2.91097089	\\
                6.644	3.84896151	\\
                6.691	5.105948941	\\
                6.738	6.214937851	\\
                6.785	7.299927001	\\
                6.832	8.232917671	\\
                6.879	8.606913931	\\
                6.926	8.688913111	\\
                6.973	8.544914551	\\
                7.02	7.797922021	\\
                7.067	6.841931581	\\
                7.114	5.610943891	\\
                7.161	4.47895521	\\
                7.208	3.45096549	\\
                7.255	2.48197518	\\
                7.302	1.78398216	\\
                7.349	1.22498775	\\
                7.396	0.74599254	\\
                7.443	0.45899541	\\
                7.49	0.26499735	\\
                7.537	0.16499835	\\
                7.584	0.09299907	\\
                7.631	0.04099959	\\
                7.678	0.02699973	\\
                7.725	0.00799992	\\
                7.772	0.00599994	\\
                7.819	0	\\
                7.866	0	\\
                7.913	0.00099999	\\
                7.96	0.00099999	\\
            };

            \addplot[ybar interval, fill=Brown1HVDC, fill opacity=0.6, draw=Brown1HVDC, area legend] table[row sep=crcr] {%    		
                x	y	
                10.9	0.00199996	\\
                10.958	0.0049999	\\
                11.016	0.00799984	\\
                11.074	0.0149997	\\
                11.132	0.0249995	\\
                11.19	0.07899842	\\
                11.248	0.1699966	\\
                11.306	0.299994	\\
                11.364	0.59198816	\\
                11.422	0.97198056	\\
                11.48	1.548969021	\\
                11.538	2.405951881	\\
                11.596	3.438931221	\\
                11.654	4.693906122	\\
                11.712	6.073878522	\\
                11.77	7.411851763	\\
                11.828	8.248835023	\\
                11.886	8.937821244	\\
                11.944	9.428811424	\\
                12.002	9.018819624	\\
                12.06	8.278834423	\\
                12.118	7.330853383	\\
                12.176	6.025879482	\\
                12.234	4.777904442	\\
                12.292	3.465930681	\\
                12.35	2.612947741	\\
                12.408	1.641967161	\\
                12.466	1.13197736	\\
                12.524	0.62398752	\\
                12.582	0.3549929	\\
                12.64	0.18199636	\\
                12.698	0.09199816	\\
                12.756	0.05799884	\\
                12.814	0.02399952	\\
                12.872	0.00599988	\\
                12.93	0.0099998	\\
                12.988	0	\\
                13.046	0.00299994	\\
                13.104	0	\\
                13.162	0.00199996	\\
                13.22	0.00199996	\\
            };

            \addplot[ybar interval, fill=Brown2HVDC, fill opacity=0.6, draw=Brown2HVDC, area legend] table[row sep=crcr] {%    		
                x	y	
                10.01	0.00399996	\\
                10.089	0.01799982	\\
                10.168	0.04399956	\\
                10.247	0.07799922	\\
                10.326	0.16799832	\\
                10.405	0.32799672	\\
                10.484	0.62899371	\\
                10.563	1.04398956	\\
                10.642	1.74898251	\\
                10.721	2.52597474	\\
                10.8	3.69596304	\\
                10.879	4.80095199	\\
                10.958	6.109938901	\\
                11.037	7.456925431	\\
                11.116	8.200917991	\\
                11.195	8.939910601	\\
                11.274	8.992910071	\\
                11.353	8.763912361	\\
                11.432	7.987920121	\\
                11.511	7.033929661	\\
                11.59	5.685943141	\\
                11.669	4.57795422	\\
                11.748	3.62396376	\\
                11.827	2.54697453	\\
                11.906	1.77098229	\\
                11.985	1.24798752	\\
                12.064	0.78599214	\\
                12.143	0.50599494	\\
                12.222	0.2899971	\\
                12.301	0.18599814	\\
                12.38	0.09299907	\\
                12.459	0.06199938	\\
                12.538	0.0299997	\\
                12.617	0.0099999	\\
                12.696	0.00299997	\\
                12.775	0.00599994	\\
                12.854	0.00099999	\\
                12.933	0	\\
                13.012	0	\\
                13.091	0.00099999	\\
                13.17	0.00099999	\\
            };

            \addplot [color=red,dashed, line width =0.6pt, forget plot] table[row sep=crcr]{%
                5.59	0	\\
                5.59   15	\\
            };
            
            \addplot [color=red,dashed, line width =0.6pt, forget plot] table[row sep=crcr]{%
                6.94	0	\\
                6.94   15	\\
            };
            
            \addplot [color=red,dashed, line width =0.6pt, forget plot] table[row sep=crcr]{%
                11.97	0	\\
                11.97   15	\\
            };
            
            \addplot [color=red,dashed, line width =0.6pt, forget plot] table[row sep=crcr]{%
                11.31	0	\\
                11.31   15	\\
            };
            \legend{Kontek, Baltic Cable, NorNed, SwePol}
        \end{axis}

    \end{tikzpicture}
    \vspace{-0.3em}
    \caption{Distribution of the average congestion rent for Kontek, Baltic Cable, NorNed and SwePol obtained after data re-sampling.}
    \label{fig:7_hvdcprices}
    \vspace{-1em}
\end{figure}
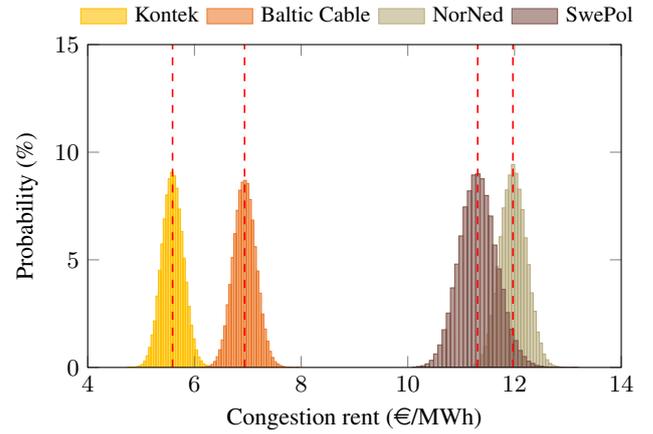

Given that the current installed capacity of synchronous condensers is not enough to make any significant difference in low inertia situations, this remedial action would require the installation of new equipment with the related costs. Alternatively, gas and hydro power units could be used as synchronous compensators. The main cost of running hydro turbines without discharging water is the cost of cooling water: for 1 GWs/h this cost would be in the range of 270-340 Euros \cite{4_4a}. However, if the turbine has no direct access to cooling water, the realization of a cooling system is necessary, with an investment cost in the range of 0.22-0.33 million Euros \cite{4_4a}. Gas turbines run at minimum power can support the system with ``real" inertia. Given that the cost of production of these units is usually high and that they often provide tertiary reserves, the average price for inertia produced by gas turbines is 3500 \euro/GWs/h \cite{2_9}.

Finally, more frequency reserve products could be procured by TSOs. In general, by reserving generation capacity there are socioeconomic costs associated with the capacity that cannot be used in the energy market. Indeed, if the generating units that provide frequency support have low fuel costs, it is likely that more expensive generators will supply the reserved power to the market. In case of new services, e.g. FFR, additional costs will be incurred for adjusting existing controllers. In case of frequency imbalance, TSOs incurs the activation cost of such reserves, that is usually higher than the spot price.
\section{Cost Saving Analyses}\label{sec:6}
\begin{figure*}[t]
    \begin{tikzpicture}
        \begin{axis}[%
            width=0.90\textwidth,
            height=0.15\textheight,
            at={(0,0)},
            scale only axis,
            axis on top,
            ymin=0,
            ymax=221,
            xmin=3900,
            xmax=5600,
            xtick={{4000},{4500},{5000},{5500}},
            xticklabels={{4'000},{4'500},{5'000},{5'500}},
            ylabel={Power (MW)},
            xlabel={Hours (h)},
            axis background/.style={white},
            xlabel near ticks,
            ylabel near ticks,
            legend columns=1,
            legend style={at={(0.95,0.96)}, anchor=north east, legend cell align=left, align=left, draw=none, font=\footnotesize},
            every axis legend/.append style={column sep=1em}
            ]
            
            \addplot [color=orange1, line width=1pt, fill=orange1, fill opacity = 0.5, area legend]
                    table {Plots/Data/DIreduction.dat};
            \addplot [color=blue1, line width=0.6pt, fill=blue1, fill opacity = 0.5, area legend]
                    table {Plots/Data/HVDCepc.dat};
            \legend{DI reduction, HVDC EPC}
                
        \end{axis}
\end{tikzpicture}
    \vspace{-0.3em}
    \caption{Comparison between DI reduction and HVDC EPC for the low inertia events of the 2025 HN scenario (low inertia).}
    \label{fig:8_occasion}
    \vspace{-1em}
\end{figure*}
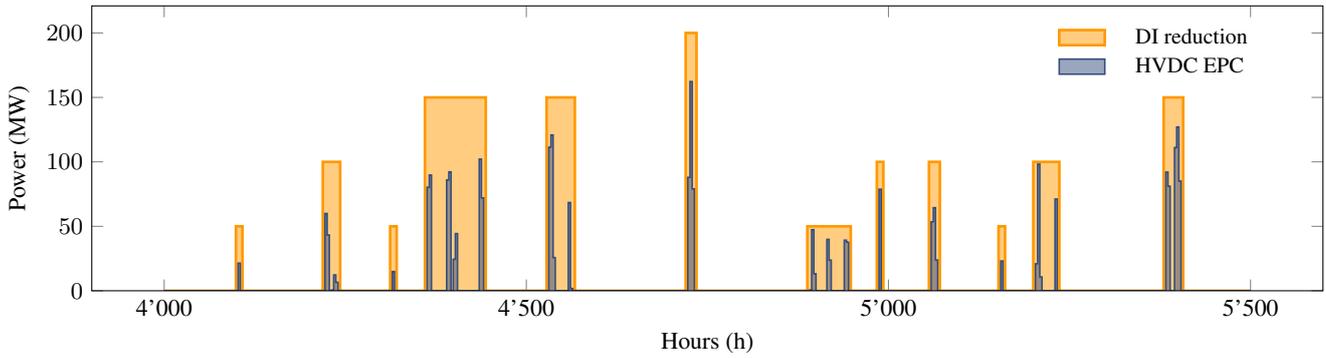

In this section, the results of the cost saving analyses are presented. The authors would like to stress that the presented results are highly dependent on the assumptions made in the previous sections, summarized as follows:

\addvbuffer[4pt 6pt]{%
\setlength{\extrarowheight}{2pt}
\hskip-1.5em
\begin{tabularx}{0.488\textwidth}{l *1{>{\arraybackslash}X}}
\vspace{-0.1em}
     $\bullet$ & The redispatch of O3 happens in the Regulating Power Market; OKG receives a fixed compensation per event (for the reduced efficiency) and a compensation proportional to the power limitation (for the opportunity costs), while the power that is not produced by O3 is procured at the regulating power price. \\
     $\bullet$ & Frequency reserves are procured in the neighboring zones and some HVDC capacity is reserved. HVDC overloading is not considered.
\end{tabularx}%
}

\begin{table}[!b]
%\vspace{-1em}
    \caption{Low inertia events and DI reduction}
    %\vspace{-0.3em}
    \label{tab2:lowperiodsDI}
    \small
    \centering
        \begin{tabularx}{0.47\textwidth}{*2{>{\arraybackslash}X} c c c}
        \hline
        \multicolumn{2}{c}{\textbf{Scenario}} & \textbf{Events}    & \textbf{Hours} (h)     & \textbf{Energy} (GWh)     \Tstrut\Bstrut\\ 
        \hline
        2018                & &  3         & 166           & 17                \Tstrut\Bstrut\\
        \multirow{3}{*}{2020} & High       & -         & -           &  -            \Tstrut\\
        & Medium       & 8         & 856           & \text{137 }             \\
        & Low       & 11         & 1'950           &  392            \Bstrut\\
        \multirow{3}{*}{2025 FN} & High       & -         & -           & -       \Tstrut\\
        & Medium       & 15         & 345           & \text{48 }             \\
        & Low       & 16         & 1'020           &   182           \Bstrut\\
        \multirow{3}{*}{2025 HN} & High       & 1         & 60           &   5           \Tstrut\\
        & Medium       & 18         & 649           & \text{132 }             \\
        & Low       & 14         & 1'638           &  437           \Bstrut\\
        \hline
    \end{tabularx}
\end{table}

\begin{table}[!b]
%\vspace{-1em}
    \caption{Costs: DI reduction (medium-low inertia, median prices)}
    %\vspace{-0.5em}
    \label{tab4:currentparadigm}
    \small
    \centering
    \begin{tabularx}{0.47\textwidth}{l *2{>{\centering\arraybackslash}X} cc}
        \hline
        & {2018} & {2020} & {2025 FN} & {2025 HN} \Tstrut\Bstrut\\
        \hline
        \multicolumn{5}{l}{\textbf{Medium inertia}}\Tstrut\\
        Down-regulation (M\euro) & 0.091             & 0.674                 & 0.294      & \text{0.698 }   \\ 
        Up-regulation (M\euro) & 0.897           & 4.740                 & 1.661        & \text{4.567 } \\
        \textbf{Total cost} (M\euro)        & \textbf{0.988}                 & \textbf{5.414}                 & \textbf{1.955}      & \textbf{5.265}      \Bstrut\\
        \hline
        \multicolumn{5}{l}{\textbf{Low inertia}}\Tstrut\\
        Down-regulation (M\euro) & 0.091             & 1.871                 & 0.920      & \text{2.094 }   \\ 
        Up-regulation (M\euro) & 0.897           & 13.563                 & 6.297        & \text{15.120 } \\
        \textbf{Total cost} (M\euro)        & \textbf{0.988}                 & \textbf{15.434}                 & \textbf{7.218}      & \textbf{17.214}      \Bstrut\\
        \hline
    \end{tabularx}
\end{table}

Thus, if one or more of the assumptions are inexact, the cost savings could be considerably larger or smaller.

\tablename~\ref{tab2:lowperiodsDI} and \ref{tab3:lowperiodsEPC} present the results of the algorithm introduced in Section \ref{sec:3}, with the hours when the kinetic energy is below the requirement and the corresponding remedial actions. The first thing to be noticed is that the occurrence of low inertia periods is highly impacted by the weather. In terms of hours when the kinetic energy is expected to be below the requirements, this number is almost always zero in case of wet hydrological years (high inertia scenarios). On the contrary, these numbers are tripled in case of dry years (compared to an average hydrological year). This happens because during dry periods the production of hydro power plants (that are synchronous machines) is replaced by imports via HVDC (which do not contribute to the kinetic energy of the system). The same trend is observed with the entity of remedial actions ("Energy" column in \tablename~\ref{tab2:lowperiodsDI} and \ref{tab3:lowperiodsEPC}), as they are proportional to the amount of hours. Conversely, the number of events does not follow the same trend: technical limitations play an important role on the length of DI reduction and, with high frequency of low inertia periods, the limitation on O3 is often prolonged resulting in less but longer events. 

\begin{table}[!b]
%\vspace{-1em}
    \caption{Low inertia events and HVDC EPC}
    %\vspace{-0.3em}
    \label{tab3:lowperiodsEPC}
    \small
    \centering
        \begin{tabularx}{0.47\textwidth}{p{1.7cm} p{1.4cm} *2{>{\centering\arraybackslash}X}}
        \hline
        \multicolumn{2}{c}{\textbf{Scenario}} & \textbf{Hours} (h)     & \textbf{Energy} (GWh)     \Tstrut\Bstrut\\ 
        \hline
        2018                &         & 166           & 19                \Tstrut\Bstrut\\
        \multirow{3}{*}{2020} & High       & -           &  -            \Tstrut\\
        & Medium       &  260           & \text{17 }             \\
        & Low       & 603           &  47            \Bstrut\\
        \multirow{3}{*}{2025 FN} & High       & -           & -       \Tstrut\\
        & Medium       & 165           & \text{11 }             \\
        & Low       & 411           &   28           \Bstrut\\
        \multirow{3}{*}{2025 HN} & High       & 12         & 1           \Tstrut\\
        & Medium       & 165           & \text{31 }             \\
        & Low       & 411           &  72           \Bstrut\\
        \hline
    \end{tabularx}
\end{table}
\begin{table}[!b]
%\vspace{-1em}
    \caption{Costs: HVDC EPC (medium-low inertia, median prices)}
    %\vspace{-0.5em}
    \label{tab5:HVDC_EPC}
    \small
    \centering
    \begin{tabularx}{0.47\textwidth}{l *2{>{\centering\arraybackslash}X} cc}
        \hline
        & {2018} & {2020} & {2025 FN} & {2025 HN} \Tstrut\Bstrut\\
        \hline
        \multicolumn{5}{l}{\textbf{Medium inertia}}\Tstrut\\
        HVDC capacity (M\euro) & 0.048             & 0.152                & 0.098      & \text{0.278 } \\
        Primary reserves (M\euro) & 0.225           & 0.172                 & 0.111        & \text{0.314 } \\
        \textbf{Total cost} (M\euro)        & \textbf{0.273}                 & \textbf{0.324}                 & \textbf{0.210}      & \textbf{0.592}      \Bstrut\\
        \hline
        \multicolumn{5}{l}{\textbf{Low inertia}}\Tstrut\\
        HVDC capacity (M\euro) & 0.048             & 0.421                & 0.251      & \text{0.645 }      \\
        Primary reserves (M\euro) & 0.225           & 0.476                 & 0.284        & \text{0.730 } \\
        \textbf{Total cost} (M\euro)        & \textbf{0.273}                 & \textbf{0.897}                 & \textbf{0.534}      & \textbf{1.374}      \Bstrut\\
        \hline
    \end{tabularx}
\end{table}

Second, the amount of hours of low inertia periods is expected to decrease in the next 5 years. This is mainly because of the new HVDC interconnections, as more transmission capacity means more exports. Indeed, the high share of hydro power makes Nordic electricity cheaper than in other countries (e.g. Netherlands or UK). Thus, more exports mean more synchronous generation. This is even reflected in the 2025 \textit{HN} scenario, where low inertia periods with half of the nuclear capacity are shorter than the ones in 2020. 

Lastly, the length of each event differs depending on the remedial action. For instance, HVDC capacity and UCTE reserves are needed only for those specific hours when the kinetic energy is below the requirements, while this is not the case for DI reduction. This can be visualized in \figurename~\ref{fig:8_occasion}, where the two remedial actions are compared.

The cost of reducing the dimensioning incident is provided in \tablename~\ref{tab4:currentparadigm}. In summer 2018, the cost of downregulating O3 is calculated to be around 90 thousand euros, while the procurement of the substitute power around 900 thousand euros, resulting in a total cost of 0.988 million euros. Depending on the type of hydrological year, future projections suggest that there could be more low-inertia periods between 2020 and 2025, resulting in considerable higher redispatch costs. Three different prices have been used for calculating the cost of procuring the substitute power, based on the historical prices from 2015 to 2019. As the standard deviation of the distribution in \figurename~\ref{fig:8_occasion} is quite small, the corresponding costs are quite close to each other (for example, in 2020 - low inertia, costs are in the range of 15.24-15.63 million euros). Thus, for the sake of space, only the results calculated with the median are displayed. On the contrary, the results vary a lot across different inertia scenarios: they are equal to or close to zero if wet years are to be expected, and triple moving from average to dry years. 

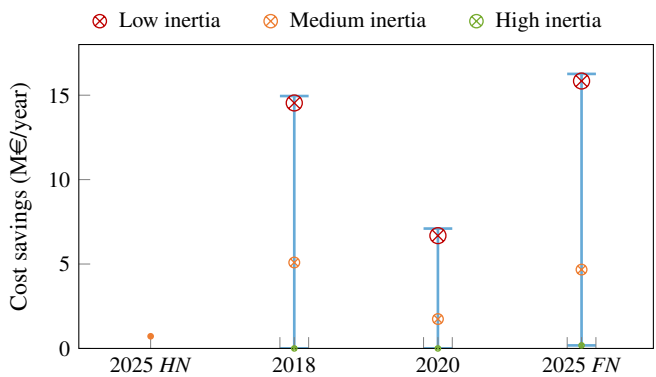
\begin{figure}
    \centering
    \begin{tikzpicture}%
        \begin{axis}[
            width  = 0.42\textwidth,
            height=0.16\textheight,
            axis on top,
            scale only axis,
            % bar shift auto,
            xmin=0.5,
            xmax=4.5,
            xtick align=inside,
            xtick pos=left,
            xticklabel style = {font=\footnotesize},
            % symbolic x coords = {2018, 2020, 2025-FN, 2025-HN},
            xtick=data,
            xticklabels={{2018}, {2020}, {2025 \textit{FN}}, {2025 \textit{HN}}},
            ymin=0,
            ymax=18,
            ytick pos=left,
            ytick={0,5,10,15},
            yticklabels={0,5,10,15},
            ylabel = {Cost savings (M\euro/year)},
            ylabel style = {font=\footnotesize},
            yticklabel style = {font=\footnotesize},
            ylabel near ticks,
            axis background/.style={fill=white},
            legend columns=3,
            legend style={at={(0.01,1.01)}, anchor=south west, legend cell align=left, align=left, draw=none, font=\footnotesize},
            /tikz/every even column/.append style={column sep=0.5cm},
            every axis legend/.append style={column sep=0.3em}
            ]
            
            \addplot[only marks, red1, mark size=3pt, mark=otimes, line width = 0.5pt, forget plot] table [row sep=crcr] {%
                2 14.54\\
                3 6.68\\
                4 15.84 \\
            };
            \addplot[only marks, OrangeHVDC, mark size=2pt, mark=otimes, line width = 0.5pt, forget plot] table [row sep=crcr] {%
                2 5.09\\
                3 1.74\\
                4 4.67 \\
            };
            \addplot[only marks, OrangeHVDC, mark size=1pt, line width = 0.5pt, forget plot] table [row sep=crcr] {%
                1 0.72\\
            };
            \addplot[only marks, GreenFCR2, mark size=1pt, mark=otimes, line width = 0.5pt, forget plot] table [row sep=crcr] {%
                2 0\\
                3 0\\
                4 0.18 \\
            };
            % 2020
            % \addplot[ybar, bar width = 10pt, draw = blue1, line width = 0.5pt, fill = blue1, fill opacity = 0.6] table [row sep=crcr] {%
            %     2 14.90\\
            % };
            % \addplot[ybar, bar width = 11pt, draw = white, fill = white] table [row sep=crcr] {%
            %     2 4.99\\
            % };
            % \addplot[blue1, line width = 0.5pt] table [row sep=crcr] {%
            %     1.902 4.99\\
            %     2.098 4.99\\
            % };
            % \addplot[only marks, blue1, mark size=1pt ] table [row sep=crcr] {%
            %     2 5.08\\
            %     2 14.54\\
            % };
            
            % 2020
            \addplot[BlueREG, line width = 1pt, forget plot] table [row sep=crcr] {%
                1.9 0\\
                2.1 0\\
            };
            \addplot[BlueREG, line width = 1pt, forget plot] table [row sep=crcr] {%
                1.9 14.95\\
                2.1 14.95\\
            };
            \addplot[BlueREG, line width = 1pt, forget plot] table [row sep=crcr] {%
                2 0\\
                2 14.95\\
            };
            
            % 2025 FN
            \addplot[BlueREG, line width = 1pt, forget plot] table [row sep=crcr] {%
                2.9 0\\
                3.1 0\\
            };
            \addplot[BlueREG, line width = 1pt, forget plot] table [row sep=crcr] {%
                2.9 7.1\\
                3.1 7.1\\
            };
            \addplot[BlueREG, line width = 1pt, forget plot] table [row sep=crcr] {%
                3 0\\
                3 7.1\\
            };
            
            % 2025 HN
            \addplot[BlueREG, line width = 1pt, forget plot] table [row sep=crcr] {%
                3.9 0.18\\
                4.1 0.18\\
            };
            \addplot[BlueREG, line width = 1pt, forget plot] table [row sep=crcr] {%
                3.9 16.26\\
                4.1 16.26\\
            };
            \addplot[BlueREG, line width = 1pt, forget plot] table [row sep=crcr] {%
                4 0\\
                4 16.26\\
            };
            
            \addplot[only marks, red1, mark size=2.5pt, mark=otimes, line width = 0.5pt] table [row sep=crcr] {%
                6 100 \\
            };
            
            \addplot[only marks, OrangeHVDC, mark size=2.5pt, mark=otimes, line width = 0.5pt] table [row sep=crcr] {%
                6 100 \\
            };
            
            \addplot[only marks, GreenFCR2, mark size=2.5pt, mark=otimes, line width = 0.5pt] table [row sep=crcr] {%
                6 100 \\
            };
            
            \legend{Low inertia, Medium inertia, High inertia}
        \end{axis}
        % \draw[help lines];
        % \draw[red1] (2.835,3.255) circle (0.075cm);
        % \draw[red1] (4.725,1.495) circle (0.065cm);
        % \draw[red1] (6.615,3.545) circle (0.085cm);
        % \draw[OrangeHVDC] (2.835,1.14) circle (0.065cm);
        % \draw[OrangeHVDC] (4.725,0.395) circle (0.065cm);
        % \draw[OrangeHVDC] (6.615,1.045) circle (0.065cm);
        % \draw[GreenFCR2] (6.615,0.04) circle (0.065cm);
    \end{tikzpicture}%
    \caption{Cost savings by using HVDC in the form of EPC for frequency support.}
    \label{fig:6_res}
    \vspace*{-0.5em}
\end{figure}

For the calculation of costs associated with HVDC EPC, it has been assumed that Nordic TSOs pays for reserving HVDC capacity and for procuring reserves on the other systems. This corresponds to the entire spectrum of costs they could bear for this action. The calculated costs are presented in \tablename~\ref{tab5:HVDC_EPC}. Also in this case, the difference between the costs calculated with high and low prices is not big, as the entity of these remedial actions is significantly smaller than the power limitations on O3. Therefore, only the results calculated with the median are displayed also in this case. In 2018, the costs of using HVDC can be divided into 48 thousand euros for reserving HVDC capacity and 225 thousand euros for procuring primary reserves in the three neighboring countries, for a total cost of 273 thousand euros. Similar to the preventive reduction of O3, these costs highly varies across inertia scenarios with the peak in the 2025 \textit{HN} - low inertia scenario. Once again, these are the maximum expected costs; real costs will depends on the type of agreement between Nordic and European TSOs.

Finally, the cost saving comparison between DI reduction and HVDC EPC is provided in \figurename~\ref{fig:6_res} for all the considered scenarios. The cost of the current paradigm is used as reference, and the savings from HVDC EPC are compared. The position of the marker show what are the cost savings in case of medium prices, while the marker size relates to the different price scenarios, e.g. to the range of variation due to high-low prices. The light blue bars show what is the range of potential savings for each year. Clearly, there is a substantial economic benefit from using HVDC for frequency support. Starting from 2018, potential cost savings amount to 0.72 million euros (72.4\%). As a large number of low inertia events is expected in 2020, the potential cost savings increase to 5.18 million euros per year (reaching 14.80 million euros per year in case of a dry year). Finally, in 2025, savings are potentially in the range of 1.70-1.78 million euros per year with the current capacity of nuclear power plants, or in the range of 4.56-4.78 million euros per year if the capacity of nuclear power plants is halved (these number will increase to 6.55-6.81 and 15.51-16.16 million euros per year in case of a dry year).

%% CONCLUSION %%%%%%%%%%%%%%%%%%%%%%%%%%%%%%%%%%%%%%%%%%%%%%%%%%%%%%%%%%%%%%%%%%%%%%%%%%%%%%%%%%%%%%%%%%%%%%%%%%%%%%
\section{Conclusion}\label{sec:7}
During summer 2018, the inertia level of the Nordic Synchronous Area (NSA) dropped below the security level three times, jeopardizing the N-1 security of the system. To deal with these situations, Svenska kraftn\"{a}t ordered the reduction of the power output of Oskarshamn 3, a nuclear power plant in Sweden, which is the most critical generating unit of the Nordic system. An estimation of the costs associated with this power limitation for the three instances in 2018 has been calculated in this paper and amount to 0.988 million euros. Given that more and more low-inertia periods are expected in the coming years, this calls for a reassessment of whether there exist more cost-efficient options which guarantee safe operation while avoiding expensive redispatching actions. 

In this paper, we have presented a decision-making support tool for comparing different mitigation strategies in case of low inertia periods. Moreover, we have investigated what is the cost of using HVDC interconnectors for the provision of frequency support, and we have performed a cost savings analysis comparing this alternative to the current paradigm. The analysis is carried out for three scenarios (2018, 2020, 2025), using historical data from Nord Pool and inertia estimations from Nordic TSOs. The cost of the remedial actions are calculated based on historical data from the past 5 years, considering the distribution of the average price obtained via data re-sampling. Our results show that, if HVDC was used in the form of Emergency Power Control, the costs in 2018 could be reduced to 0.27 million euro. The extension of the analysis to year 2020 and 2025 confirms that many more low-inertia periods can be expected in the future, calling for more redispatching actions. In this regard, the method proposed in this paper could reduce the costs by 70\%, resulting in cost savings in the range of 1.70-4.78 million euros per year by 2025 (or even higher if dry years are to be expected).

Although the focus of the paper is on the utilization of HVDC lines for frequency support, other mitigation strategies can be compared following the same approach. For instance, a new product - Fast Frequency Reserves (FRR) - will become available for Nordic TSOs starting from summer 2020. According to the availability of market data, the proposed tool can be used to compare the new product to the already analyzed measures.

%% ACKNOWLEDGMENTS %%%%%%%%%%%%%%%%%%%%%%%%%%%%%%%%%%%%%%%%%%%%%%%%%%%%%%%%%%%%%%%%%%%%%%%%%%%%%%%%%%%%%%%%%%%%%%%%%%%%%%
\section{Acknowledgments}
This work is supported by the multiDC project, funded by Innovation Fund Denmark, Grant Agreement No. 6154-00020B.

% APPENDIX %%%%%%%%%%%%%%%%%%%%%%%%%%%%%%%%%%%%%%%%%%%%%%%%%%%%%%%%%%%%%%%%%%%%%%%%%%%%%%%%%%%%%%%%%%%%%%%%%%%%%%%%%
% \appendices
% \input{09_appendix.tex}

%% BIBLIOGRAPHY %%%%%%%%%%%%%%%%%%%%%%%%%%%%%%%%%%%%%%%%%%%%%%%%%%%%%%%%%%%%%%%%%%%%%%%%%%%%%%%%%%%%%%%%%%%%%%%%%%%

% \bibliographystyle{myIEEEtran.bst}
% \bibliography{CS.bib}{}

\end{document}